\documentclass[pra,twocolumn,superscriptaddress,shownopacs,amssymb]{revtex4-2}
\usepackage{graphicx,psfrag,amsmath,amssymb}

\usepackage[usenames, dvipsnames]{color}

\usepackage{hyperref} 
\hypersetup{
    colorlinks=true,
    linkcolor=Blue,
    citecolor=Blue,
    filecolor=Blue,      
    urlcolor=Blue,
    pdftitle={Few-fermions},
    }
    
\begin{document}

\title{Stability of $(N+1)$-body fermion clusters in multiband Hubbard model}

\author{M. Iskin}
\affiliation{
Department of Physics, Ko\c{c} University, Rumelifeneri Yolu, 34450 Sar\i yer, Istanbul, Turkey
}

\author{A. Kele\c{s}}
\affiliation{
Department of Physics, Middle East Technical University, Ankara, 06800, Turkey
}

\date{\today}

\begin{abstract}

We start with a variational approach and derive a set of coupled integral equations 
for the bound states of $N$ identical spin-$\uparrow$ fermions and a single 
spin-$\downarrow$ fermion in a generic multiband Hubbard Hamiltonian 
with an attractive onsite interaction. As an illustration we apply our integral equations 
to the one-dimensional sawtooth lattice up to $N \le 3$, i.e., to the $(3+1)$-body problem, 
and reveal not only the presence of tetramer states in this two-band model but also 
their quasi-flat dispersion when formed in a flat band. Furthermore, 
for $N = \{4, 5, \cdots, 10 \}$, our DMRG simulations and exact diagonalization suggest 
the presence of larger and larger multimers with lower and lower binding energies, 
conceivably without an upper bound on $N$. These peculiar $(N+1)$-body clusters are in 
sharp contrast with the exact results on the single-band linear-chain model where none 
of the $N \ge 2$ multimers appear. Hence their presence must be taken into account for 
a proper description of the many-body phenomena in flat-band systems, e.g., they may 
suppress superconductivity especially when there exists a large spin imbalance.

\end{abstract}

\maketitle

\section{Introduction}
\label{sec:intro}

Exactly solvable few-body problems offer a unique avenue to gain valuable insights 
into the microscopic origins of novel many-body phenomena, starting from a collection 
of isolated composites such as dimers, trimers, and other 
multimers~\cite{braaten06, hammer10, blume12, greene17, naidon17, dincao18, mattis86}. 
The archetypal example is the instability of a non-interacting Fermi gas against the 
formation of Cooper pairs, which eventually leads the way to the theory of BCS 
superconductivity. It turns out a dimer that is formed between a spin up and a 
spin down fermion (assuming they have equal masses) is the only stable bound state 
in the presence of an attractive contact interaction, suggesting that the underlying 
Cooper-pairing mechanism is robust for the BCS theory in all dimensions~\cite{mattis86}. 
In sharp contrast with this conventional insight, here we show that the multimers 
may also play a decisive role in a multiband system, especially when there exists 
a flat band in the spectrum.  

Until recently all of the few-body studies were focused either on continuum systems 
or on their lattice counterparts that feature only a single band. As an example three 
identical bosonic atoms that are interacting via short-range resonant interactions 
in vacuum is known to exhibit an infinite series of three-body bound states, i.e., 
the so-called Efimov effect~\cite{braaten06, hammer10, blume12, greene17, naidon17, dincao18, mattis86}. 
The experimental realization of this long-sought trimer state with ultracold 
bosons~\cite{kraemer06, zaccanti09, pollack09, gross09, grimm19} have in return
sparked a growing interest in related effects with fermions. For instance three-body, 
four-body and and five-body Efimov effects have all been predicted, respectively, 
with two, three and four identical heavy fermions that are interacting resonantly 
with a much lighter particle~\cite{kartavtsev07, castin10, levinsen13, jag14, 
blume12b, bazak17, sanayei20, liu22}.
Thus, unlike the equal-mass case where only dimer states are allowed, 
the multimer (trimer, tetramer and pentamer) states appear in mass-imbalanced 
mixtures when the mass-ratio exceeds a certain threshold depending on the 
multimer type. See recent reviews for a comprehensive list of related 
works from atomic, molecular and optical physics to condensed-matter, nuclear 
and particle physics~\cite{braaten06, hammer10, blume12, greene17, naidon17, dincao18}.
Analogous predictions were also reported for the appearance of 
trimer states in single-band lattices but only when the tunneling amplitudes are 
spin dependent~\cite{orso10, orso11, roux11, dalmonte12, dhar18}. 

Despite all these progress it is surprising that the few-body physics is still in 
its infancy in a more realistic lattice model when there exists more than one 
Bloch band in the one-body spectrum. For instance, as opposed to the exact results 
on the single-band linear-chain which allow only dimers~\cite{takahashi70, 
mattis86, orso10, orso11}, the energetic stability of the trimers have recently 
been predicted in a sawtooth lattice which features two bands~\cite{orso22, iskin22b}. 
Remarkably these trimers have a quasi-flat dispersion with a negligible bandwidth 
when they form in a flat band, which is very different from the 
highly-dispersive spectrum of the underlying dimers. 
In this paper we study the bound states of $N$ identical spin-$\uparrow$ 
fermions and a single spin-$\downarrow$ fermion in a generic multiband Hubbard 
model with an attractive onsite interaction. We start with a variaional approach, 
and derive a set of exact results that are readily applicable to all lattice geometries 
in all dimensions. As an illustration we apply our $N \le 3$ theory to the sawtooth 
lattice, and reveal both the energetic stability of the tetramer states and their 
quasi-flat dispersion when formed in a flat band. 
Furthermore we perform DMRG simulations and exact diagonalization to investigate 
the possibility of larger bound states, and present strong evidence for the energetic 
stability of the multimer states with $N = \{4, 5, \cdots, 10\}$, conceivably 
without an upper bound on $N$.

Given the recent surge of experimental and theoretical interest in flat-band 
systems~\cite{jo12, nakata12, li18, diebel16, kajiwara16, ozawa17, slot17, huda20,
arovas21,parameswaran13, liu14, leykam18, balents20}, which is boosted by the 
discovery of superconductivity and correlated insulating states in the magic-angle 
twisted bilayer graphene (MATBG) systems~\cite{cao18}, we hope that our peculiar 
findings will trigger further interest in the few-body aspects of Kagome and Lieb-like 
toy models that exhibit a flat band in their spectrum~\cite{mizoguchi19}. 
Furthermore one of the profound implications of our findings is that the Cooper-pairing 
based theories of superconductivity in flat-band systems may not always be the best 
starting point, i.e., the formation of multimers may suppress superconductivity 
in these systems, especially when there exists a large spin imbalance. Thus our 
exact few-body results will shed some light on the proper description of the many-body 
phenomena in flat-band systems. 

The rest of this paper is organized as follows. In Sec.~\ref{sec:Np1} we first introduce
the multiband Hubbard model and then derive the integral equations for the $(N+1)$-body 
problem through a variational approach.
In Sec.~\ref{sec:sawtooth} we apply our variational results to the sawtooth lattice, 
and discuss the full $(1+1)$-body spectrum, full $(2+1)$-body spectrum, and ground 
state of tetramers. There we also analyze ground states of pentamers and other 
multimers with DMRG simulations and exact diagonalization. 
In Sec.~\ref{sec:conc} we end the paper with a brief summary of our results and outlook. 
As for the Appendices, the numerical implementation of the $(3+1)$-body problem is 
described in App.~\ref{sec:ni}, the low-energy excitation energies from the exact 
diagonalization are presented in App.~\ref{sec:low}, and the fermion-boson mapping 
in the three-body problem is illustrated in App.~\ref{sec:fbm}.

\section{$(N+1)$-Body Problem}
\label{sec:Np1}

In this paper we are interested in the role of multiple Bloch bands on the energetic 
stability of the multimer states. For this purpose we consider few-body bound states 
that are made of $N$ spin-$\uparrow$ fermions and a single spin-$\downarrow$ fermion
in a multiband Hubbard model.

\subsection{Multiband Hubbard model}
\label{sec:model}

The standard Hubbard Hamiltonian~\cite{arovas21, qin21}
$
\mathcal{H} = \sum_\sigma \mathcal{H}_\sigma + \mathcal{H}_{\uparrow\downarrow}
$
is made of two terms:
\begin{align}
\label{eqn:H_sigma}
\mathcal{H}_\sigma &= -\sum_{Si; S'i'} t_{Si; S'i'}^\sigma c_{S i \sigma}^\dagger c_{S' i' \sigma},
\\
\label{eqn:H_updo}
\mathcal{H}_{\uparrow\downarrow} &= - U \sum_{S i} 
c_{S i \uparrow}^\dagger c_{S i \downarrow}^\dagger 
c_{S i \downarrow} c_{S i \uparrow}.
\end{align}
Here the first term accounts for the kinetic energy of the spin-$\sigma$ fermions where 
the hopping parameter $t_{Si; S'i'}^\sigma$ describes their tunneling amplitude from 
the sublattice (or basis or orbital) site $S'$ in the unit cell $i'$ to the sublattice site $S$ 
in the unit cell $i$. On the other hand the second term accounts for the potential energy 
of the system, where $U \ge 0$ is the strength of the attractive interaction between 
$\uparrow$ and $\downarrow$ fermions when they are on the same site. 
Next we use a canonical transformation, i.e., 
\begin{align}
c_{S i \sigma}^\dagger = \frac{1}{\sqrt{N_c}} \sum_\mathbf{k} 
e^{-i \mathbf{k} \cdot \mathbf{r_{S i}}} c_{S \mathbf{k} \sigma}^\dagger,
\end{align}
and express the Hubbard Hamiltonian in the reciprocal lattice, where $N_c$ is the 
number of unit cells in the system, $\mathbf{k}$ is the crystal momentum in the 
first Brillouin zone (BZ), and $\mathbf{r_{S i}}$ is the position of the sublattice 
site $S$ in unit cell $i$. 
Note that the total number of lattice sites is $N_s = N_b N_c$ when the number of 
sublattice sites in a unit cell is $N_b$.
In addition noting that
$
c_{n \mathbf{k} \sigma}^\dagger = \sum_S n_{S \mathbf{k} \sigma} 
c_{S \mathbf{k} \sigma}^\dagger,
$
where $n$ is the band index for the Bloch bands (there are $N_b$ of them) and 
$n_{S \mathbf{k} \sigma}$ is the projection of the Bloch state onto the sublattice $S$, 
we eventually find~\cite{iskin21}
\begin{align}
\label{eqn:Hsigma}
\mathcal{H}_\sigma &= \sum_{n \mathbf{k}} \varepsilon_{n\mathbf{k}\sigma}
c_{n \mathbf{k} \sigma}^\dagger c_{n \mathbf{k} \sigma},
\\
\label{eqn:Hupdown}
\mathcal{H}_{\uparrow\downarrow} &= \frac{1}{N_c} 
\sum_{\substack{nmn'm' \\ \mathbf{k}\mathbf{k'}\mathbf{q}}}
V_{n'm'\mathbf{k'}}^{nm\mathbf{k}}(\mathbf{q})
b_{nm}^\dagger(\mathbf{k}, \mathbf{q})
b_{n'm'}(\mathbf{k'}, \mathbf{q}).
\end{align}
Here $\varepsilon_{n\mathbf{k}\sigma}$ is the one-body dispersion of the fermions 
in band $n$,
$
V_{n'm'\mathbf{k'}}^{nm\mathbf{k}}(\mathbf{q}) = - U\sum_S 
n_{S, \mathbf{k}+\frac{\mathbf{q}}{2}, \uparrow}^*
m_{S, -\mathbf{k}+\frac{\mathbf{q}}{2}, \downarrow}^*
{m'}_{S, -\mathbf{k'}+\frac{\mathbf{q}}{2}, \downarrow}
{n'}_{S, \mathbf{k'}+\frac{\mathbf{q}}{2}, \uparrow}
$
characterizes the onsite interactions in momentum space, and
$
b_{nm}^\dagger (\mathbf{k}, \mathbf{q}) = c_{n,\mathbf{k}+\frac{\mathbf{q}}{2}, \uparrow}^\dagger
c_{m,-\mathbf{k}+\frac{\mathbf{q}}{2}, \downarrow}^\dagger
$
creates a pair of fermions in the Bloch bands.

\subsection{Variational approach}
\label{sec:va}

Motivated by the success of variational approach on the two-body and 
three-body problems~\cite{iskin21, iskin22, iskin22b}, 
here we attack the $(N+1)$-body problem with the following ansatz
\begin{align}
\label{eqn:HPsi}
\mathcal{H} | \Psi_\mathbf{q} \rangle &= E_{N+1}^\mathbf{q} | \Psi_\mathbf{q} \rangle,
\\
\label{eqn:Psiq}
| \Psi_\mathbf{q} \rangle =& 
\sum_{\substack{n_1\cdots n_N m \\ \mathbf{k_1} \cdots \mathbf{k_N}}} 
\alpha_{n_1 \cdots n_N m}^{\mathbf{k_1} \cdots \mathbf{k_N}} (\mathbf{q})
\bigg( \prod_{i=1}^N c_{n_i \mathbf{k_i} \uparrow}^\dagger \bigg)
c_{m \mathbf{Q} \downarrow}^\dagger
| 0 \rangle.
\end{align}
Here the ansatz $| \Psi_\mathbf{q} \rangle$ explicitly conserves the center-of-mass (CoM) 
momentum $\mathbf{q}$ of the particles, $E_{N+1}^\mathbf{q}$ is the energy of the 
($N+1$)-body bound state, 
$
\alpha_{n_1 \cdots n_N m}^{\mathbf{k_1} \cdots \mathbf{k_N}} (\mathbf{q})
$
is the variational complex parameter, and
\begin{align}
\mathbf{Q} = \mathbf{q} - \sum_{i=1}^{N}  \mathbf{k_i}
\end{align}
is defined for convenience. Since $| \Psi_\mathbf{q} \rangle$ has the most general form, 
it will provide us the exact solution. 
The normalization condition can be written as
$
\langle \Psi_\mathbf{q} | \Psi_\mathbf{q} \rangle = 
N! \sum_{\substack{n_1 \cdots n_N m \\ \mathbf{k_1} \cdots \mathbf{k_N}}} 
|\alpha_{n_1 \cdots n_N m}^{\mathbf{k_1} \cdots \mathbf{k_N}} (\mathbf{q})|^2,
$
where we enforce the Pauli principle, and make extensive use of
\begin{align}
\alpha_{n_1 \cdots n_i \cdots n_j \cdots n_N m}^{\mathbf{k_1} \cdots \mathbf{k_i} 
\cdots \mathbf{k_j} \cdots \mathbf{k_N}} (\mathbf{q})
= - \alpha_{n_1 \cdots n_j \cdots n_i \cdots n_N m}^{\mathbf{k_1} \cdots \mathbf{k_j} 
\cdots \mathbf{k_i} \cdots \mathbf{k_N}} (\mathbf{q}),
\end{align}
i.e., the ansatz picks up a minus sign under the exchange of a pair of its 
spin-$\uparrow$ fermions. After a lengthy but a straightforward calculation, 
we find
\begin{align}
\langle \mathcal{H}_\uparrow \rangle &= 
N! \sum_{\substack{n_1 \cdots n_N m \\ \mathbf{k_1} \cdots \mathbf{k_N}}} 
|\alpha_{n_1 \cdots n_N m}^{\mathbf{k_1} \cdots \mathbf{k_N}} (\mathbf{q})|^2 
\bigg( \sum_{i = 1}^N \varepsilon_{n_i \mathbf{k_i} \uparrow} \bigg),
\\
\langle \mathcal{H}_\downarrow \rangle &= 
N! \sum_{\substack{n_1 \cdots n_N m \\ \mathbf{k_1} \cdots \mathbf{k_N}}} 
|\alpha_{n_1 \cdots n_N m}^{\mathbf{k_1} \cdots \mathbf{k_N}} (\mathbf{q})|^2 
\varepsilon_{m \mathbf{Q} \downarrow},
\\
\langle \mathcal{H}_{\uparrow\downarrow} \rangle &= 
-\frac{N! U}{N_c} 
\sum_{\substack{n_1 \cdots n_N m_1 m_2 n \\ S \mathbf{k_1} \cdots \mathbf{k_N} \mathbf{k}}} 
\alpha_{n_1 \cdots n_N m_1}^{\mathbf{k_1} \cdots \mathbf{k_N}} (\mathbf{q})
{n}^*_{S \mathbf{k} \uparrow} {m_1}_{S \mathbf{Q} \downarrow}
\nonumber \\
\times \sum_{i=1}^{N} & \bigg\lbrace 
 \sum_{n_i \mathbf{k_i}} 
[\alpha_{n_1 \cdots n_N m_2}^{\mathbf{k_1} \cdots \mathbf{k_N}} (\mathbf{q})]^*
{m_2}^*_{S \mathbf{Q} \downarrow} \delta_{n_i n} \delta_{\mathbf{k_i} \mathbf{k}}
\bigg\rbrace 
{n_i}_{S \mathbf{k_i} \uparrow},
\end{align}
for the expectation value of the multiband Hubbard Hamiltonian. Here $^*$ is for the
complex conjugation and $\delta_{ij}$ is the Kronecker delta.

The variational parameters are determined through the functional minimization of 
$
\langle \Psi_\mathbf{q} | \mathcal{H} - E_{N+1}^\mathbf{q} | \Psi_\mathbf{q} \rangle,
$
but this procedure leads to a complicated expression. In order to simplify the resultant 
equations, we define a new parameter set
\begin{align}
\gamma_{n_2 \cdots n_N S}^{\mathbf{k_2} \cdots \mathbf{k_N}}(\mathbf{q}) 
= \sum_{n_1 m \mathbf{k_1}}
\alpha_{n_1 \cdots n_N m}^{\mathbf{k_1} \cdots \mathbf{k_N}}(\mathbf{q})
{n_1}_{S \mathbf{k_1} \uparrow} {m}_{S \mathbf{Q} \downarrow},
\end{align}
and make use of Pauli exchange statistics
$
\gamma_{n_2 \cdots n_i \cdots n_j \cdots n_N S}^{\mathbf{k_2} \cdots \mathbf{k_i} 
\cdots \mathbf{k_j} \cdots \mathbf{k_N}} (\mathbf{q}) 
= - 
\gamma_{n_2 \cdots n_j \cdots n_i \cdots n_N S}^{\mathbf{k_2} \cdots \mathbf{k_j} 
\cdots \mathbf{k_i} \cdots \mathbf{k_N}} (\mathbf{q}).
$
We finally obtain a set of coupled integral equations with the following structure 
\begin{widetext}
\begin{align}
\gamma_{n_2 \cdots n_N S}^{\mathbf{k_2} \cdots \mathbf{k_N}} (\mathbf{q}) 
&= \frac{U}{N_c} \sum_{n_1 m S' \mathbf{k_1}}
\frac{{m}_{S' \mathbf{Q} \downarrow}^* {m}_{S \mathbf{Q} \downarrow} {n_1}_{S \mathbf{k_1} \uparrow}}
{\big( \sum_{i = 1}^N \varepsilon_{n_i \mathbf{k_i} \uparrow} \big) 
+ \varepsilon_{m \mathbf{Q} \downarrow} - E_{N+1}^\mathbf{q}}
\nonumber \\
& \times
\bigg\lbrace
{n_1}_{S' \mathbf{k_1} \uparrow}^* 
\gamma_{n_2 \cdots n_N S'}^{\mathbf{k_2} \cdots \mathbf{k_N}}(\mathbf{q})
- \sum_{i = 2}^{N}  {n_i}_{S' \mathbf{k_i} \uparrow}^* 
\bigg[ \sum_{n_i \mathbf{k_i}}
\gamma_{n_2 \cdots n_N S'}^{\mathbf{k_2} \cdots\mathbf{k_N}}(\mathbf{q})
\delta_{n_i n_1} \delta_{\mathbf{k_i} \mathbf{k_1}}
\bigg]
\bigg\rbrace.
\label{eqn:gammaN}
\end{align}
\end{widetext}
This exact expression is one of our central results in this work: the ($N+1$)-body 
problem in a multiband Hubbard model is reduced to the solutions of $N_b^N$ 
coupled integral equations with $N-1$ momentum variables for a given set of 
parameters, i.e., $\mathbf{q}$, $U$ and hoppings. 
Its continuum version is recovered by setting the Bloch factors to unity and 
dropping the band as well as sublattice indices, i.e., it requires the solution of a single 
integral equation for $\gamma^{\mathbf{k_2} \cdots\mathbf{k_N}}(\mathbf{q})$,
see Eq.~(3) in Ref.~\cite{liu22}, and Ref.~\cite{pricoupenko11} for details.
Once $E_{N+1}^\mathbf{q}$ is obtained, the binding energy of the ($N+1$)-body bound 
state can be determined by
\begin{align}
\label{eqn:E_bind}
E_{N+1}^\textrm{be} (\mathbf{q}) = -E_{N+1}^\mathbf{q} 
+ \min \{ E_{(N-1)+1}^\mathbf{q'} + \varepsilon_{n, \mathbf{q} - \mathbf{q'}, \uparrow} \}.
\end{align}
This is because while an ($N+1$)-body bound state may in general become 
energetically unstable against dissociation into an $[(N-\ell)+1]$-body bound state 
and $\ell$ free spin-$\uparrow$ fermions, the $\ell = 1$ process is closest in 
energy to $E_{N+1}^\mathbf{q}$ when the $[(N-1)+1]$-body bound state is 
energetically-stable, i.e., $E_N^\textrm{be} (\mathbf{q}) > 0$, to begin 
with~\footnote{It is such that the binding energy $E_{2}^\textrm{be} (\mathbf{q})$ 
of the dimer is always defined from an unbound pair of a free spin-$\downarrow$ 
fermion plus a free spin-$\uparrow$ fermion; 
the binding energy $E_{3}^\textrm{be} (\mathbf{q})$ of the trimer is defined 
from the dimer threshold plus a free spin-$\uparrow$ fermion 
when $E_{2}^\textrm{be} (\mathbf{q}) > 0$; 
the binding energy $E_{4}^\textrm{be} (\mathbf{q})$ of the tetramer is defined 
from the trimer threshold plus a free spin-$\uparrow$ fermion 
when $E_{3}^\textrm{be} (\mathbf{q}) > 0$, etc.}.
Indeed this turns out to be the case for all of the multimers in the flat-band 
of sawtooth lattice discussed below.

Let's first show that Eq.~(\ref{eqn:gammaN}) reproduces the available literature
in the $N = 1$ and $N = 2$ cases.
For $N = 1$, since the summation term of the second line is irrelevant, 
Eq.~(\ref{eqn:gammaN}) is equivalent to
\begin{align}
\gamma_{S} (\mathbf{q}) 
&= \frac{U}{N_c} \sum_{n_1 m S' \mathbf{k_1}}
\frac{m_{S' \mathbf{Q} \downarrow}^* m_{S \mathbf{Q} \downarrow} {n_1}_{S \mathbf{k_1} \uparrow}{n_1}_{S' \mathbf{k_1} \uparrow}^*}
{\varepsilon_{n_1 \mathbf{k_1} \uparrow} + \varepsilon_{m \mathbf{Q} \downarrow} - E_2^\mathbf{q}}
\gamma_{S'} (\mathbf{q}),
\label{eqn:gamma1}
\end{align}
where $\mathbf{Q} = \mathbf{q} - \mathbf{k_1}$. This self-consistency relation can be
recast as an eigenvalue problem of an $N_b \times N_b$ matrix, giving rise to $N_b$ 
branches for the two-body dispersion $E_2^\mathbf{q}$ for each given $\mathbf{q}$. 
See Ref.~\cite{orso22} for an alternative derivation with a different approach.
Equation~(\ref{eqn:gamma1}) is recently used to reveal a deeper connection 
between the effective-mass tensor of the lowest-lying dimer states and the 
quantum-metric tensor of the underlying Bloch states~\cite{iskin21, iskin22, torma18}.
For $N = 2$ Eq.~(\ref{eqn:gammaN}) reduces to
\begin{align}
\gamma_{n_2 S}^{\mathbf{k_2}} (\mathbf{q}) 
&= \frac{U}{N_c} \sum_{n_1 m S' \mathbf{k_1}}
\frac{m_{S' \mathbf{Q} \downarrow}^* m_{S \mathbf{Q} \downarrow} {n_1}_{S \mathbf{k_1} \uparrow}}
{\varepsilon_{n_1 \mathbf{k_1} \uparrow}  + \varepsilon_{n_2 \mathbf{k_2} \uparrow} + \varepsilon_{m \mathbf{Q} \downarrow} - E_3^\mathbf{q}}
\nonumber \\
 &\times \big[
{n_1}_{S' \mathbf{k_1} \uparrow}^* 
\gamma_{n_2 S'}^{\mathbf{k_2}} (\mathbf{q})
- {n_2}_{S' \mathbf{k_2} \uparrow}^* \gamma_{n_1 S'}^{\mathbf{k_1}} (\mathbf{q})
\big],
\label{eqn:gamma2}
\end{align}
where $\mathbf{Q} = \mathbf{q} - \mathbf{k_1} - \mathbf{k_2}$. This is a set of $N_b^2$ 
coupled integral equations with one momentum variable, and it can be recast as an 
eigenvalue problem of an $N_b^2 N_c \times N_b^2 N_c$ matrix for each given 
$\mathbf{q}$. Equation~(\ref{eqn:gamma2}) has recently been derived by one of
us~\cite{iskin22b}, 
and its numerical solutions for $E_3^\textrm{be} (\mathbf{q})$ are in excellent 
agreement with the DMRG simulations in a sawtooth lattice~\cite{orso22}. 
In particular, in sharp contrast with the exact results on the single-band linear-chain 
model which dismiss trimers~\cite{takahashi70, mattis86, orso10, orso11}, 
it is found that the presence of an additional band allows the formation of
energetically-stable trimer states in the sawtooth lattice. 
In addition it is found that the trimers have a quasi-flat dispersion when formed 
in a flat band, which is unlike the highly-dispersive spectrum of its dimers. 
These surprising results are one of the main motivations for the current work, 
i.e., we want to study the stability of larger few-body clusters in the 
presence of multiple bands.

Let's next consider the four-body problem and study the fate of tetramer bound states. 
For $N = 3$ Eq.~(\ref{eqn:gammaN}) reduces to
\begin{align}
\gamma_{n_2 n_3 S}^{\mathbf{k_2} \mathbf{k_3}} (\mathbf{q}) 
&= \frac{U}{N_c} \sum_{n_1 m S' \mathbf{k_1}}
\frac{m_{S' \mathbf{Q} \downarrow}^* m_{S \mathbf{Q} \downarrow} {n_1}_{S \mathbf{k_1} \uparrow}}
{\big( \sum_{i = 1}^3 \varepsilon_{n_i \mathbf{k_i} \uparrow} \big) + \varepsilon_{m \mathbf{Q} \downarrow} - E_4^\mathbf{q}}
\nonumber \\
 & \times \big[
{n_1}_{S' \mathbf{k_1} \uparrow}^* 
\gamma_{n_2 n_3 S'}^{\mathbf{k_2} \mathbf{k_3}} (\mathbf{q})
- {n_2}_{S' \mathbf{k_2} \uparrow}^* \gamma_{n_1 n_3 S'}^{\mathbf{k_1} \mathbf{k_3}} (\mathbf{q})\nonumber \\
& \quad \quad  \quad \quad  \quad \quad \quad
- {n_3}_{S' \mathbf{k_3} \uparrow}^* \gamma_{n_2 n_1 S'}^{\mathbf{k_2} \mathbf{k_1}} (\mathbf{q})
\big],
\label{eqn:gamma3}
\end{align}
where $\mathbf{Q} = \mathbf{q} - \mathbf{k_1} - \mathbf{k_2} - \mathbf{k_3}$.
This is a set of $N_b^3$ coupled integral equations with two momentum variables, 
and it can be recast as an eigenvalue problem of an $N_b^3 N_c^2 \times N_b^3 N_c^2$ 
matrix for each given $\mathbf{q}$. 
Our numerical recipe is provided in App.~\ref{sec:ni}. 
In this work we apply Eqs.~(\ref{eqn:gamma1}),~(\ref{eqn:gamma2}) and~(\ref{eqn:gamma3}) 
to the sawtooth model due in part to its flat band and one-dimensional simplicity, 
and most importantly to our benchmarking capacity with the DMRG simulations and 
exact diagonalization.

\section{Sawtooth Lattice}
\label{sec:sawtooth}

Due to the presence of its $N_b = 2$ sublattice sites in a unit cell (say $S = \{A, B\}$
sublattices), the sawtooth lattice features two Bloch bands in the first BZ 
(say $s = \{+, -\}$ bands). See the inset of Fig.~\ref{fig:variational}(a) 
for its sketch where $a$ is the lattice spacing. 
Here we allow hopping between nearest-neighbor sites only, and set 
$
t_{Aj;Ai}^\sigma = - t
$
with $j = i \pm 1$ and $t \ge 0$,
$
t_{Bj;Bi}^\sigma = 0
$
and 
$
t_{Bi;Ai}^\sigma = t_{Bj;Ai}^\sigma = - t'
$
with $j = i-1$ and $t' \ge 0$. Thus the one-body Hamiltonian can be written as
\begin{align}
\mathcal{H}_\sigma = \sum_k \psi_{k \sigma}^\dagger
\big(d_k^0 \sigma_0 + \mathbf{d}_k \cdot \boldsymbol{\sigma} \big)
\psi_{k \sigma},
\end{align}
where
$
\psi_{k \sigma} = ( c_{A k \sigma} \,\, c_{B k \sigma} )^\mathrm{T}
$
is a sublattice spinor, $-\pi/a < k \le \pi/a$ is in the first BZ, 
$
d_k^0 = t \cos(k a),
$
$\sigma_0$ is a $2\times2$ identity matrix,
$
\mathbf{d}_k = (d_k^x, d_k^y, d_k^z)
$
is a field vector with elements
$
d_k^x = t' + t' \cos(k a),
$
$
d_k^y = t' \sin(k a)
$
and 
$
d_k^z = t \cos(k a),
$
and
$
\boldsymbol{\sigma}  = (\sigma_x, \sigma_y, \sigma_z)
$
is a vector of Pauli spin matrices. The one-body dispersions can be written as
$
\varepsilon_{s k \sigma} =  d_k^0 + s d_k
$
where $s = \pm$ for the upper and lower bands, respectively, and $d_k$ is the 
magnitude of $\mathbf{d}_k$. 
The sublattice projections of the corresponding eigenvectors are
$
s_{A k \sigma} = (- d_k^x + id_k^y)/\sqrt{2d_k(d_k - sd_k^z)}
$
and
$
s_{B k \sigma} = (d_k^z - sd_k) / \sqrt{2d_k(d_k - sd_k^z)}.
$
One of the most treasured features of this toy model is the presence of a 
flat (lower) band $\varepsilon_{-, k} = -2t$ in its dispersion when 
$t'/t = \sqrt{2}$~\cite{huber10, phillips15, pyykkonen21, 
chan22, orso22, iskin22b}.

\subsection{Full $(1+1)$-body spectrum}
\label{sec:1p1}

Equation~(\ref{eqn:gamma1}) determines only the two lowest-energy dimer
bound states~\cite{iskin21, iskin22, iskin22b, orso22}. 
In order to reveal the full two-body spectrum, one may recast Eq.~(\ref{eqn:gamma1}) 
as an eigenvalue problem in terms of the original variational parameters,
\begin{align}
\label{eqn:gamma2SE}
&0 = (\varepsilon_{n_1 \mathbf{k_1} \uparrow} + \varepsilon_{n_2, \mathbf{q - k_1}, \downarrow} 
- E_2^\mathbf{q} )
\alpha_{n_1 n_2}^{\mathbf{k_1}} (\mathbf{q}) 
\\
&-
 \frac{U}{N_c} \sum_{n_1' n_2' \mathbf{k} S}
{n_1^*}_{S \mathbf{k_1} \uparrow} {n_2^*}_{S, \mathbf{q - k_1}, \downarrow} 
{n_2'}_{S, \mathbf{q-k}, \downarrow} {n_1'}_{S \mathbf{k} \uparrow} 
\alpha_{n_1' n_2'}^{\mathbf{k}} (\mathbf{q})
\nonumber,
\end{align}
and solve for $E_2^\mathbf{q}$~\cite{iskin21}. This can be achieved by further recasting 
it as an eigenvalue problem using an $N_b^2 N_c \times N_b^2 N_c$ matrix for each 
given $\mathbf{q}$. Here we use a $k$-space mesh with $N_c = 100$ points 
for the two-body problem.

\begin{figure*}[!htb]
    \centering
    \includegraphics[width=1.98\columnwidth]{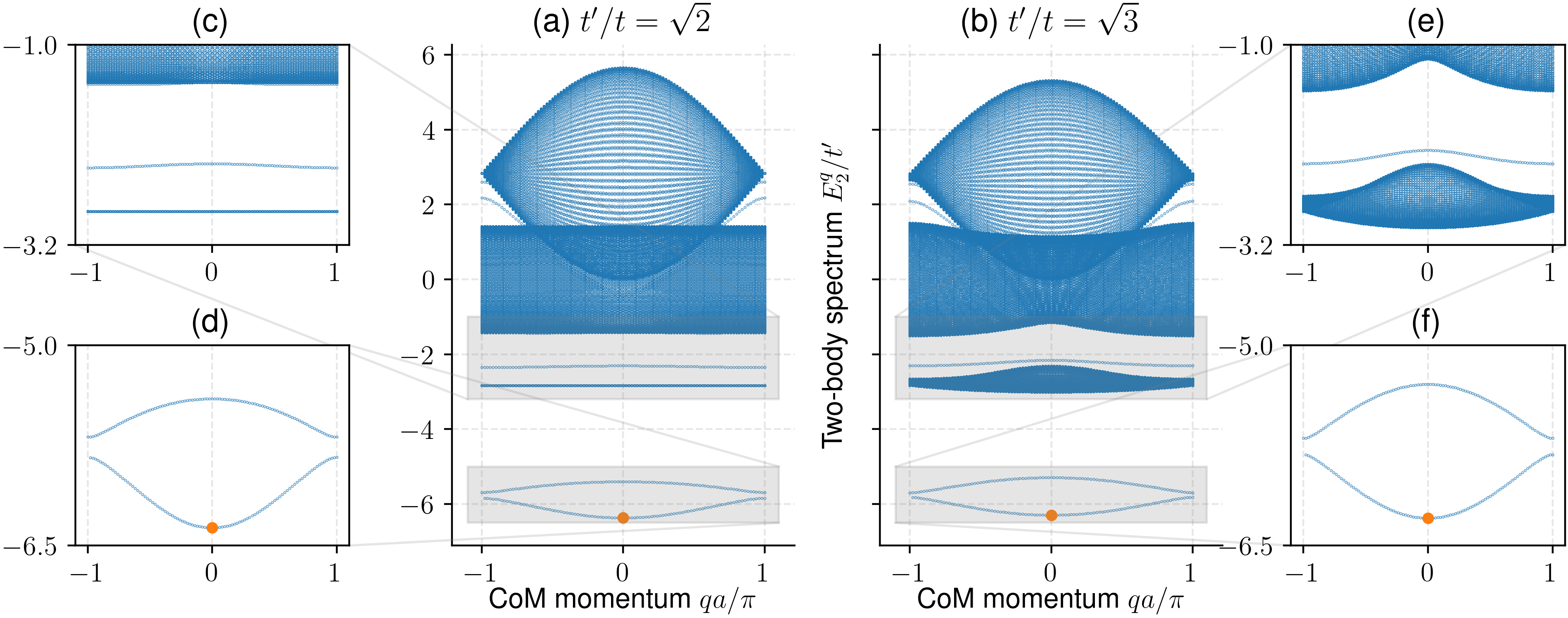}
    \caption{Two-body spectum $E_2^q$ in a sawtooth lattice for 
    $t = t'/\sqrt{2}$ (left column) and $t = t'/\sqrt{3}$ (right column),
    when $U = 5t'$. 
    Insets (d) and (f) are zooms to the lower and upper onsite dimer branches.
    Insets (c) and (e) are zooms to the first and second monomer-monomer continua, 
    and two distinct offsite dimer branches in between. 
    In (c) the highly-degenerate first continuum states appear precisely at 
    $-2\sqrt{2}t' \approx -2.83t'$, and the lower offsite dimer branch is around $-2.35t'$. 
    The upper offsite dimer branch is underneath the second continuum 
    but it is not visible in this scale.     
    Benchmarks with the DMRG ground states are shown with orange-colored marker $\bullet$.
    } 
    \label{fig:fulldim} 
\end{figure*} 

Our two-body spectra are shown in Fig.~\ref{fig:fulldim} for $t'/t = \{\sqrt{2}, \sqrt{3}\}$
when $U = 5t'$. At the bottom of each spectrum, there are two distinct bound states 
for a given $q$. We refer to them as the onsite dimer states (they are also referred to 
as doublons~\cite{orso22}) since their binding energy grows with $U$ without a limit, 
i.e., the two monomers are eventually tightly bound and they are strongly 
co-localized on one site in the strong-coupling limit. 
They are shown in Figs.~\ref{fig:fulldim}(d) and~\ref{fig:fulldim}(f).
Above the onsite dimers, there are the first monomer-monomer continuum states,
corresponding to two unbound monomers that occupy the lower Bloch band. 
Note that there is a continuum of highly-degenerate states at energy 
$-2\sqrt{2}t' \approx -2.83t'$ in the flat-band case shown 
in Fig.~\ref{fig:fulldim}(c), i.e., when $t'/t = \sqrt{2}$.  
The next monomer-monomer continuum states correspond to two unbound monomers 
that occupy the lower and upper Bloch bands one at a time. 
In between the first and the second continuum there are also the so-called 
offsite dimer bound states. There are two of them but the upper one 
is barely visible since it is very close to the bottom of the second continuum. 
Unlike those of the onsite dimers, the binding energy of the offsite dimers 
saturates with $U$, i.e., the two monomers are weakly bound no matter what $U$ is. 
The closest the two monomers can get is on nearest-neighbor sites in the 
strong-coupling limit. 
The third monomer-monomer continuum states correspond to two unbound monomers 
that occupy the upper Bloch band. In between the second and the third continuum 
there also appears two weakly-bound offsite dimer bound states near the
edges of the BZ. 
Thus we conclude that the offsite dimer states are a multiband phenomenon, 
and they emerge in between every two consecutive monomer-monomer continua.

We note that the presence of a flat Bloch band does not have much impact on the 
dispersion of the onsite dimers. This is because the diverging bare effective 
band mass of the flat-band monomers gets dressed by the interband transitions, 
and produce a finite effective band mass for the onsite dimers at any 
$U \ne 0$. This somewhat counter-intuitive effect is well-studied in the
recent literature in the context of quantum geometry~\cite{iskin21, iskin22}. 
See also Ref.~\cite{orso22} for the sawtooth lattice.
The effective band mass of the onsite dimers makes a dip in the weak-coupling 
regime, and then it increases for stronger couplings, leading to a more and 
more localized onsite dimers in space. This is because an onsite dimer is 
allowed to hop in the Hubbard model through the so-called virtual ionization, 
and this brings a factor of $1/E_2^\textrm{be}(\mathbf{0})$ as a punishment 
from second-order perturbation theory.
On the other hand the offsite dimers have a negligible dispersion when they 
form in a flat band, e.g., in Fig.~\ref{fig:fulldim}(c). This is because 
their effective band mass is largely controlled by the bare effective band 
mass of the weakly-bound monomers.

\subsection{Full $(2+1)$-body spectrum}
\label{sec:2p1}

Equation~(\ref{eqn:gamma2}) can be used to calculate the energy of the ground-state 
trimers through an iterative procedure~\cite{iskin22b}. However, in order to reveal 
the full three-body spectrum, we recast Eq.~(\ref{eqn:gamma2}) as an eigenvalue 
problem in terms of the original variational parameters,
\begin{align}
\label{eqn:gamma3SE}
&0 = (\varepsilon_{n_1 \mathbf{k_1} \uparrow} + \varepsilon_{n_2 \mathbf{k_2} \uparrow} + \varepsilon_{n_3 \mathbf{Q} \downarrow} 
- E_3^\mathbf{q} )
\alpha_{n_1 n_2 n_3}^{\mathbf{k_1} \mathbf{k_2}} (\mathbf{q})
\\
&-\frac{U}{2N_c} \sum_{n_1' n_3' \mathbf{k} S}
{n_1^*}_{S \mathbf{k_1} \uparrow} {n_3^*}_{S \mathbf{Q} \downarrow}
{n_3'}_{S, \mathbf{q-k_2-k},\downarrow } {n_1'}_{S \mathbf{k} \uparrow}
\alpha_{n_1' n_2 n_3'}^{\mathbf{k} \mathbf{k_2}} (\mathbf{q}) 
\nonumber \\
&- \frac{U}{2N_c} \sum_{n_2' n_3' \mathbf{k} S}
{n_2^*}_{S \mathbf{k_2} \uparrow} {n_3^*}_{S \mathbf{Q} \downarrow}
{n_3'}_{S, \mathbf{q-k_1-k},\downarrow } {n_2'}_{S \mathbf{k} \uparrow}
\alpha_{n_1 n_2' n_3'}^{\mathbf{k_1} \mathbf{k}} (\mathbf{q}) 
\nonumber \\
&+ \frac{U}{2N_c} \sum_{n_2' n_3' \mathbf{k} S}
{n_1^*}_{S \mathbf{k_1} \uparrow} {n_3^*}_{S \mathbf{Q} \downarrow}
{n_3'}_{S, \mathbf{q-k_2-k},\downarrow } {n_2'}_{S \mathbf{k} \uparrow}
\alpha_{n_2 n_2' n_3'}^{\mathbf{k_2} \mathbf{k}} (\mathbf{q})
\nonumber \\
&+ \frac{U}{2N_c} \sum_{n_1' n_3' \mathbf{k} S}
{n_2^*}_{S \mathbf{k_2} \uparrow} {n_3^*}_{S \mathbf{Q} \downarrow}
{n_3'}_{S, \mathbf{q-k_1-k},\downarrow } {n_1'}_{S \mathbf{k} \uparrow}
\alpha_{n_1' n_1 n_3'}^{\mathbf{k} \mathbf{k_1}} (\mathbf{q})
\nonumber,
\end{align}
and solve for $E_3^\mathbf{q}$. Recall that $\mathbf{Q} = \mathbf{q - k_1 - k_2}$
when $N = 2$, and note that the exchange-symmetry constraints are imposed on 
the variational parameters by construction.
This can be achieved by further recasting it as an eigenvalue problem using an 
$N_b^3 N_c^2 \times N_b^3 N_c^2$ matrix for each given $\mathbf{q}$. 
Here we use a $k$-space mesh with $N_c = 50$ points for the three-body problem.
The numerical procedure is similar to the one given in App.~\ref{sec:ni}.

\begin{figure*}[!htb]
    \centering
    \includegraphics[width=1.98\columnwidth]{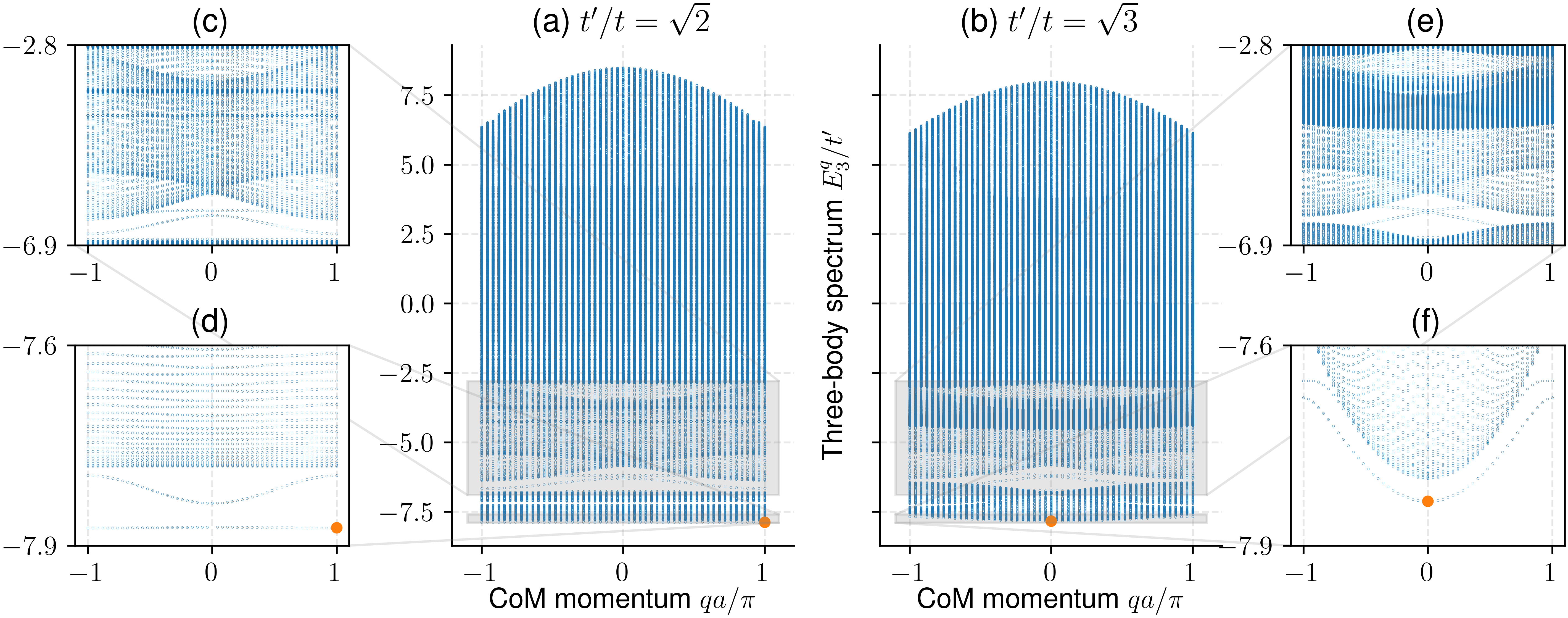} 
    \caption{Three-body spectum $E_3^q$ in a sawtooth lattice for 
    $t = t'/\sqrt{2}$ (left column) and $t = t'/\sqrt{3}$ (right column)
    when $U = 5t'$.
    Insets (d) and (f) are zooms to the offsite trimer branches, and the 
    first and second dimer-monomer continua above them.
    Insets (c) and (e) are zooms to the excited offsite trimer branches, and 
    the third and fourth dimer-monomer continua above them. 
    In (c) there is also a highly-degenerate first monomer-monomer-monomer continuum 
    states appearing precisely at $-3\sqrt{2}t' \approx -4.24t'$, and the origin of the 
    continuum of states around $-2.35t' - \sqrt{2}t' \approx -3.76t'$ can be traced 
    back to the lower offsite dimer branch shown in Fig.~\ref{fig:fulldim}(c). 
    Benchmarks with the DMRG ground states are shown with orange-colored markers $\bullet$.
    } 
    \label{fig:fulltrim} 
\end{figure*} 

Our three-body spectra are shown in Fig.~\ref{fig:fulltrim} for $t'/t = \{\sqrt{2}, \sqrt{3}\}$
when $U = 5t'$. See also Fig.~\ref{fig:figbf} in App.~\ref{sec:fbm} for similar results 
when $U = 10t'$, where some of the features discussed below are more visible. 
At the bottom of each spectrum, there are typically two distinct 
bound-state solutions for a given $q$. Here we refer to both of them as the offsite 
trimer states since their binding energies again saturate with $U$. 
They are shown in Figs.~\ref{fig:fulltrim}(d) and~\ref{fig:fulltrim}(f).
The lower trimer branch is in perfect agreement with the recent 
literature~\cite{orso22, iskin22b}.
These trimers consist of a dimer that is strongly localized on one site and a monomer 
on another site even in the strong-coupling limit, i.e., the closest the dimer and 
the monomer can be is on nearest-neighbor sites. 
See also our remarks in Sec.~\ref{sec:pentamers} for their binding mechanism.
Note that the trimer bound states are necessarily offsite and they are weakly-bound 
due to the Pauli exclusion principle preventing the formation of onsite trimers. 
Above the offsite trimers, there are the first dimer-monomer continuum 
states, corresponding to two monomers occupying 
the lower onsite dimer branch and a monomer occupying the lower Bloch band. 
The next dimer-monomer continuum states correspond to two monomers in the upper 
onsite dimer branch and a monomer in the lower Bloch band.
In the flat-band case we note that the energy gap between these dimer-monomer 
continua corresponds exactly to the energy gap between the onsite dimer branches
shown in Fig.~\ref{fig:fulldim}(d).
Within this energy gap there also appears a barely visible offsite trimer 
bound state beneath the second continuum. 
The third dimer-monomer continuum states correspond to two monomers in the 
lower onsite dimer branch and a monomer in the upper Bloch band, and 
the fourth dimer-monomer continuum to two monomers in the upper onsite dimer 
branch and a monomer in the upper Bloch band.
In between the second and the third dimer-monomer continuum there appears 
three weakly-bound offsite trimer states. 
They are shown in Figs.~\ref{fig:fulltrim}(c) and~\ref{fig:fulltrim}(e).
All of the monomer-monomer-monomer continuum states, corresponding to three 
unbound monomers occupying various combinations of the upper and lower Bloch 
bands, appear mixed together. For instance there is a continuum of highly-degenerate 
states at energy $-3\sqrt{2}t' \approx -4.24t'$ in the flat-band case shown in
Fig.~\ref{fig:fulltrim}(c), which comes from three monomers occupying the 
lower Bloch band. The next continuum that consists of two monomers in the 
lower Bloch band and a monomer in the upper Bloch band clearly starts 
at $-2\sqrt{2}t' \approx -2.83t'$.
The origin of the continuum of states around $-2.35t' - \sqrt{2}t' \approx -3.76t'$ 
can be traced back to the occupation of the lower offsite dimer branch 
shown in Fig.~\ref{fig:fulldim}(c) by two monomers along with a monomer 
in the lower Bloch band.

\subsection{Ground-state tetramers}
\label{sec:tetramers}

\begin{figure}[htb]
    \centering
    \includegraphics[width=0.99\columnwidth]{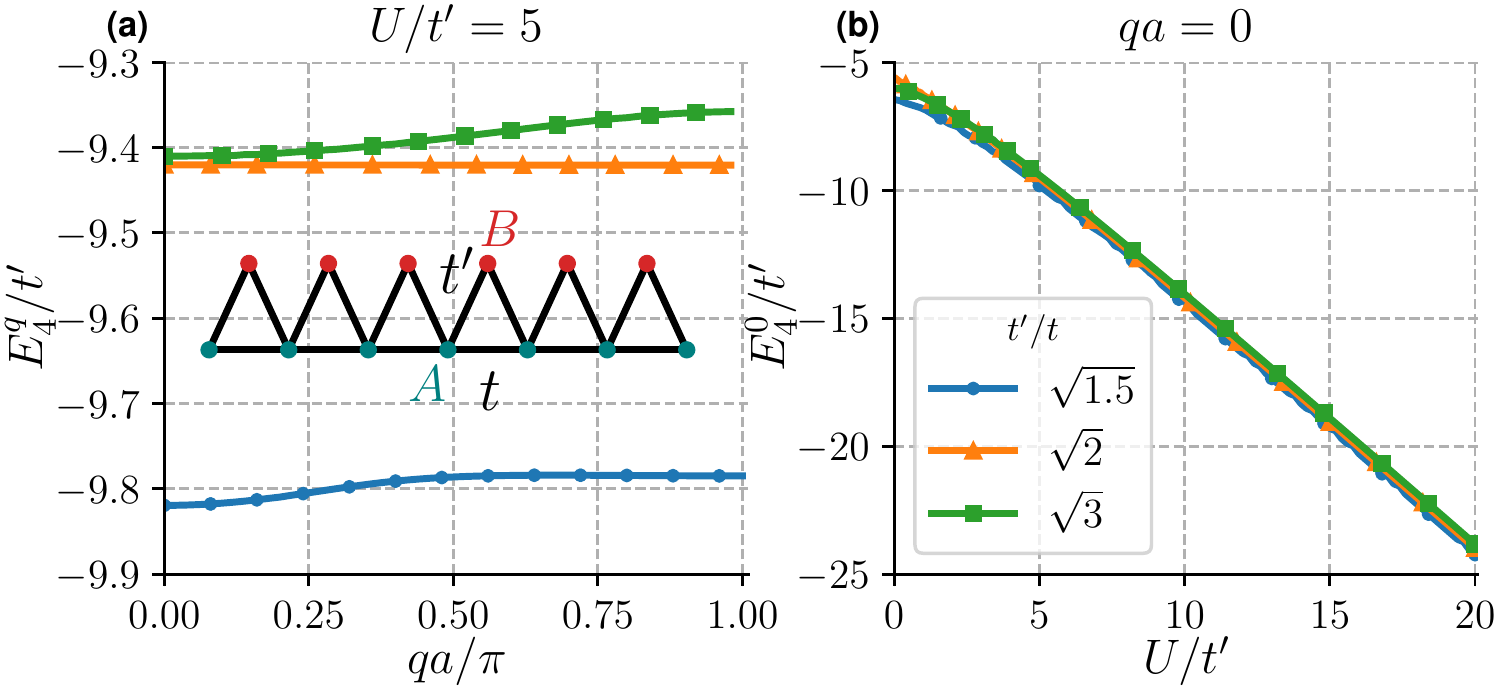}
    \caption{Lowest-lying tetramer energies from the variational approach. 
    In (a) the tetramer dispersion is quasi-flat when they form in a flat band, 
    i.e., when $t'/t = \sqrt{2}$. Note that the sawtooth lattice is sketched 
    in the inset of (a).
    } 
    \label{fig:variational}
\end{figure}

Here we analyze the ground state of the $(3+1)$-body problem. 
See App.~\ref{sec:ni} for its numerical implementation. 
Our variational results for the $E_{4}^q$ and $E_{4}^0$ are presented, respectively, 
in Figs.~\ref{fig:variational}(a) and~\ref{fig:variational}(b), where we use a $k$-space 
mesh with $N_c = 30$ points and checked that using $N_c = 50$ points makes minor corrections. 
Indeed the ground-state energy of the tetramers is typically within $1\%$ 
relative accuracy with the DMRG simulations (see below). One of our main findings is that 
the four-body dispersion $E_{4}^q$ is quasi-flat (with a negligible bandwidth) 
when the tetramers form in a flat band, i.e., when $t'/t = \sqrt{2}$. 
For instance $U = 5t'$ case is shown in Fig.~\ref{fig:variational}(a), 
and we found similar results for lower and higher $U/t'$ values as well (not shown). 
It is conceivable that the tetramers have a respectable dispersion in the weak-coupling 
limit when $U/t' \lesssim 1$, but our numerical calculations are not expected to be 
as reliable there. This is because one needs to use a much higher $N_c$ as the size 
of the bound states (in real space) gets much larger in the $U/t' \to 0$ limit. 
We also calculated the binding energy $E_{4}^\textrm{be}(q)$ of the tetramers 
and verified their energetic stability: e.g., we found that $E_{4}^\textrm{be}(q)$ 
becomes positive as soon as $U \ne 0$ when the tetramers form in a flat band. 
However this is not the case when $t'/t \ne \sqrt{2}$, i.e., $E_{4}^\textrm{be}(q)$ 
becomes positive beyond a critical threshold on $U$ in such a way that larger deviations 
from the flat-band case leads to a higher threshold.

\subsection{Ground-state pentamers and beyond}
\label{sec:pentamers}

For $N \ge 4$ it is possible to solve Eq.~(\ref{eqn:gammaN}) again by recasting it as 
an eigenvalue problem, but such a numerically-expensive task is beyond our capacity. 
Instead here we present our numerical results from the DMRG 
simulations~\cite{srwhite, schollwock11, itensor} 
and exact diagonalization~\cite{ed}. For this purpose we define the ground-state 
binding energy of the $(N+1)$-body bound state as 
\begin{align}
\label{eqn:E_bind_dmrg}
E_{N+1}^\textrm{be}(\textrm{gs}) = - E_0(N, 1) + E_0(N-1,1) + E_0(1,0),
\end{align}
where $E_0(N_\uparrow, N_\downarrow)$ is the ground-state energy of the 
$(N_\uparrow+N_\downarrow)$-body problem. Given the definition in Eq.~(\ref{eqn:E_bind}), 
Eq.~(\ref{eqn:E_bind_dmrg}) is strictly valid under the assumption that the 
CoM momentum of the ground-state of the $(N+1)$-body problem is equal
to the total momentum of the ground-states of the $[(N-1)+1]$-body and one-body problems. 
Unlike the $t'/t < \sqrt{2}$ case where the ground-state of the one-body problem is 
at the edge of the BZ, our variational results suggest that this requirement is usually 
fulfilled when $t'/t \ge \sqrt{2}$. 

\begin{figure}[htb]
    \centering
    \includegraphics[width=0.99\columnwidth]{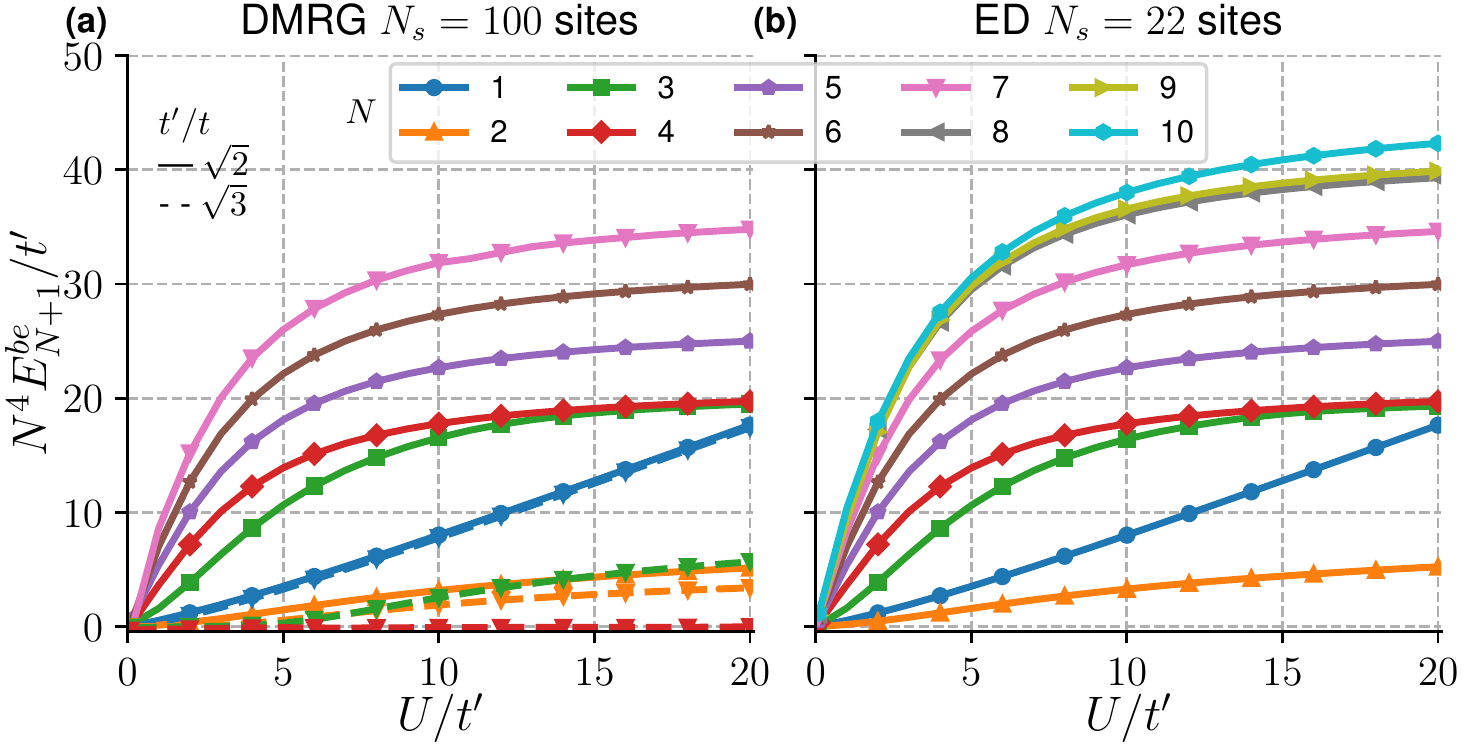}
    \caption{(a) Binding energies $E_{N+1}^\textrm{be}(\textrm{gs})$  
    from DMRG with $N_s=100$ sites for $t'/t=\sqrt{2}$ (solid lines) and 
    $t'/t=\sqrt{3}$ (dashed lines). When $N \ge 5$, the multimers are not energetically 
    stable for $t'/t = \sqrt{3}$, i.e., $E_{N+1}^\textrm{be}(\textrm{gs}) \le 0$ (not shown). 
    (b) $E_{N+1}^\textrm{be}(\textrm{gs})$ 
    from exact diagonalization with $N_s = 22$ sites for $t'/t = \sqrt{2}$. 
    Since $E_{N+1}^\textrm{be}(\textrm{gs})$ decays rapidly with $N$, 
    energies are multiplied with $N^4$ in both figures for convenience.}
    \label{fig:dmrg_ed}
\end{figure}

\begin{figure*}[!htb]
    \centering
    \includegraphics[width=1.98\columnwidth]{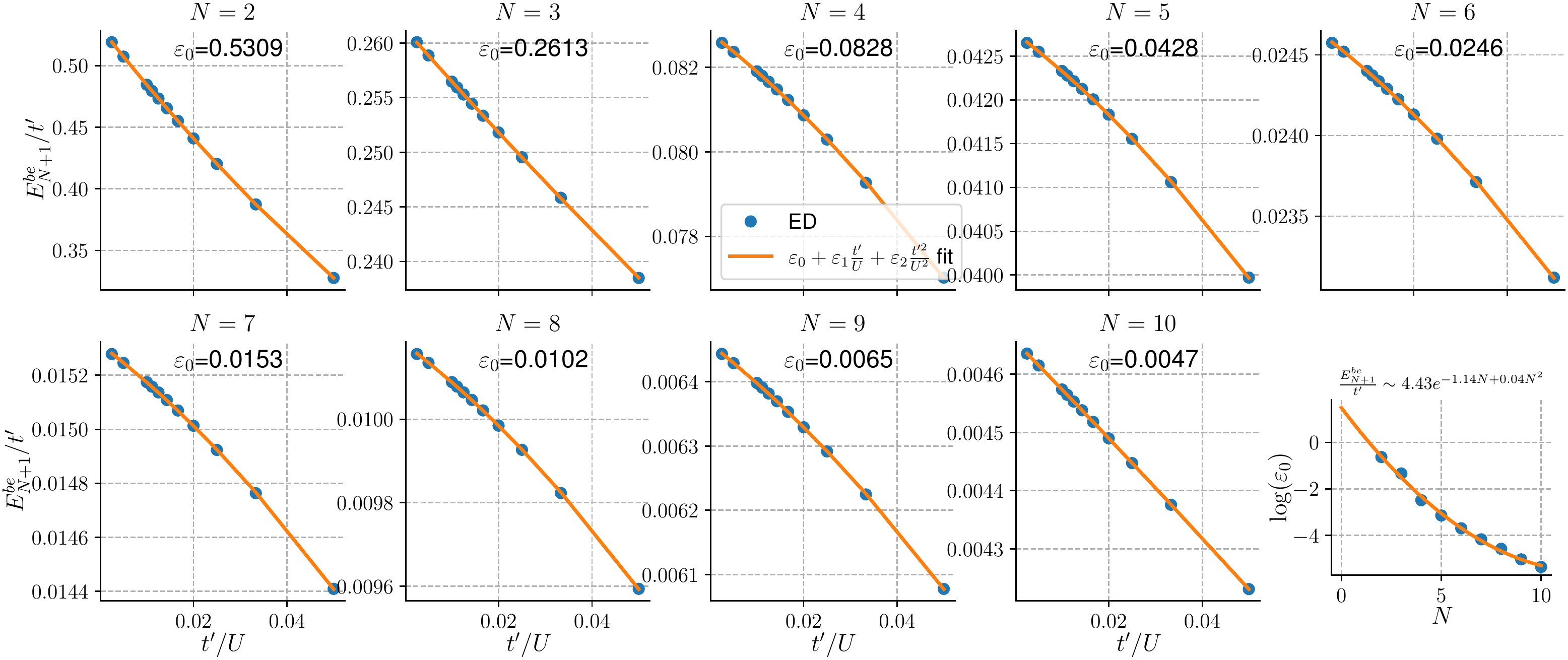}
    \caption{Binding energies $E_{N+1}^\textrm{be}(\textrm{gs})$ from exact diagonalization 
    with $N_s=22$ sites for $t'/t=\sqrt{2}$. 
    The strong-coupling limit $t'/U \ll 1$ seems to fit quite well with 
    $
    E_{N+1}^\textrm{be}(\textrm{gs})/t' = \varepsilon_0 +\varepsilon_1 (t'/U) + \varepsilon_2 (t'/U)^2
    $
    for all $N \ge 2$, where the saturation point $\varepsilon_0$ at $t'/U \to 0$ decreases 
    rapidly with $N$. It is also shown that the fit 
    $
    \varepsilon_0 \sim 4.43 e^{-1.14 N + 0.04N^2}
    $ 
    matches reasonably well with the available data for $N = \{2, 3, \cdots, 10\}$.
    }
    \label{fig:ed_sc}
\end{figure*}

In Fig.~\ref{fig:dmrg_ed}(a) we set $t'/t = \{\sqrt{2}, \sqrt{3} \}$, and present the 
DMRG results for $E_{N+1}^\textrm{be}(\textrm{gs})$ as a function of $U/t'$. 
Here we use a long lattice with $N_s = 100$ sites and with open boundary conditions. 
We only show $N = \{1, 2, \cdots, 7\}$ since the accuracy of our DMRG simulations 
does not allow us to resolve $E_{N+1}^\textrm{be}(\textrm{gs})$ for the entire $U/t'$ 
range when $N \ge 8$.
To overcome this limitation, we also perform the exact diagonalization of a fairly large 
lattice with $N_s = 22$ sites, and they are presented in Fig.~\ref{fig:dmrg_ed}(b) 
for $N = \{1, 2, \cdots, 10\}$.
First of all the variational, DMRG and exact diagonalization approaches are 
in very good agreement with each other when they have an overlap at low $N$ values. 
For $N \ge 2$ they suggest the presence of larger and larger few-body clusters with 
lower and lower binding energies, conceivably without an upper bound on $N$.
In addition all of these clusters are energetically stable when formed in a flat band, 
i.e., $E_{N+1}^\textrm{be}(\textrm{gs}) > 0$ as soon as $U \ne 0$. 
Unlike $E_{2}^\textrm{be}(\textrm{gs})$ of the dimer that grows linearly with $U$ in the 
strong-coupling limit when $U/t' \gg 1$, we note that $E_{N+1}^\textrm{be}(\textrm{gs})$ 
always saturates for $N \ge 2$, i.e., it fits quite well with $C_N' t' - C_N'' t'^2/U$
where $C_N'$ and $C_N''$ both decay rapidly with $N$. 
These fits are shown in Fig.~\ref{fig:ed_sc}.
In addition we also checked the energies of the first few excited states in our exact 
diagonalization studies. As shown in Fig.~\ref{fig:ed_gaps} in App.~\ref{sec:low}, 
the energy gaps between the first few excited states and the ground 
state vanish exactly when $t'/t = \sqrt{2}$. Thus it is also conceivable that some 
of the lowest-lying $(N+1)$-body bound states have quasi-flat dispersions in the BZ 
when formed in a flat band.

A plausible mechanism for the appearence of energetically-stable $(N+1)$-body bound states 
in a flat band is as follows: when the spin-$\uparrow$ fermions are strongly localized 
on the non-overlapping localized states (thanks to their diverging effective band mass), 
the delocalization of the spin-$\downarrow$ fermion on these nearest-neighbor states might 
be favored by the onsite attraction in between. 
See App.~\ref{sec:low} for a related numerical observation and more about localized states. 
This mechanism is analogous to the band 
formation for a single particle in a periodic potential, and the mobility of 
spin-$\downarrow$ fermion can be achieved through interband transitions 
with the upper band, reducing the overall energy of the system.
We note that a similar interband mechanism (that is mediated by $U \ne 0$) is fully 
responsible for the finite effective-mass of the low-energy dimers when they 
form in a flat band~\cite{iskin21, iskin22, orso22, torma18}.
In this scenario the spin-$\downarrow$ fermion can hop between spin-$\uparrow$ 
fermions only through dissociation of the virtual dimers, and this process leads to 
an effective hopping that scales as $\sim C_N t'^2/E_2^\textrm{be}(\textrm{gs})$ where $C_N$ 
depends on $N$. This further suggests that 
$
E_{N+1}^\textrm{be}(\textrm{\textrm{gs}}) \sim C_N't' - C_N'' t'^2/U
$ 
in the strong-coupling limit when $U/t' \gg 1$ because $E_2^\textrm{be}(\textrm{gs}) \to U$ 
in this limit. Indeed this turns out to be the case in our exact diagonalization 
results that are shown in Fig.~\ref{fig:ed_sc}. 

In the particular case of trimer bound states, this binding mechanism is in 
accordance with the one revealed for the bosonic trimers (i.e., mediated by a 
particle-exchange interaction between the onsite dimer and the monomer)~\cite{valiente10}, 
and has recently been studied through a perturbation theory in the strong-coupling 
limit~\cite{orso22}.
Furthermore, given the fermion-boson mapping that is revealed in App.~\ref{sec:fbm}, 
there is no doubt that the offsite fermion trimers are also bound through the very 
same mechanism in the strong-coupling limit.

\section{Conclusion}
\label{sec:conc}

In summary here we used variational, DMRG and exact diagonalization approaches, and 
studied the bound states of $N$ identical spin-$\uparrow$ fermions and a single 
spin-$\downarrow$ fermion in a generic multiband Hubbard Hamiltonian with an 
attractive onsite interaction. In the case of a sawtooth lattice with a flat band, 
we showed strong evidence for the existence of energetically-stable 
few-body clusters with $N = \{2, 3, \cdots,10\}$, conceivably without an upper bound 
on $N$ and with a quasi-flat $(N+1)$-body dispersion in the first BZ. 
These peculiar findings are in sharp contrast with the exact results on the 
single-band linear-chain model which dismiss all of the $N \ge 2$ 
multimers~\cite{takahashi70, mattis86, orso10, orso11}. 
As an outlook it is desirable to perform similar analyses for other 
toy models that exhibit flat bands in their spectrum including, 
e.g., the Kagome and Lieb lattices~\cite{mizoguchi19}. Such analyses would help 
us uncover how versatile the critical role of a flat band is in the formation of 
few-body clusters, and how the induced attraction between the dimer and the rest 
of the fermions~\cite{jag14} arise through interband processes. In addition 
one can easily extend our approach to study bound-state formation in a 
multiband Bose-Hubbard model~\cite{iskin22d}.

\begin{acknowledgments}
We thank M. \"O. Oktel for his comments and suggestions.
A.K. is supported by  T{\"U}B{\.I}TAK 2236 Co-funded Brain Circulation Scheme 2 
(CoCirculation2) Project No. 120C066.
\end{acknowledgments}

\appendix

\section{Numerical implementation of the $(3+1)$-body problem}
\label{sec:ni}

Equation~(\ref{eqn:gamma2}) is a set of $N_b^3$ coupled integral equations with two 
momentum variables, and here we show how to recast it as an eigenvalue problem using
an $N_b^3 N_c^2 \times N_b^3 N_c^2$ matrix for each given $\mathbf{q}$. The level 
of difficulty is the same as in the full spectrum of the $(2+1)$-body problem 
discussed in Sec.~\ref{sec:2p1}. We note that the full spectrum of the $(3+1)$-body 
spectrum is well beyond our moderate computation capacity. 
First we rewrite Eq.~(\ref{eqn:gamma2}) as
\begin{align}
\label{eqn:gamma2N}
\gamma_{n m S}^{\mathbf{k} \mathbf{k'}} (\mathbf{q}) &= 
\sum_{S'} f_{nmS; nmS'}^{\mathbf{q} \mathbf{k} \mathbf{k'}} 
\gamma_{n m S'}^{\mathbf{k} \mathbf{k'}} (\mathbf{q}) 
\nonumber\\
&+ \sum_{n' m' S' \mathbf{p}} g_{nmS; n'm'S'}^{\mathbf{q} \mathbf{k} \mathbf{k'} \mathbf{p}} 
\gamma_{n'm S'}^{\mathbf{p} \mathbf{k'}} (\mathbf{q})
\nonumber\\
&+ \sum_{n' m' S' \mathbf{p}} h_{nmS; n'm'S'}^{\mathbf{q} \mathbf{k} \mathbf{k'} \mathbf{p}} 
\gamma_{nn' S'}^{\mathbf{k} \mathbf{p}} (\mathbf{q}),
\end{align}
whose coefficients $f_{nmS; nmS'}^{\mathbf{q} \mathbf{k} \mathbf{k'}}$, 
$g_{nmS; n'm'S'}^{\mathbf{q} \mathbf{k} \mathbf{k'} \mathbf{p}}$ and
$h_{nmS; n'm'S'}^{\mathbf{q} \mathbf{k} \mathbf{k'} \mathbf{p}}$ 
are stored as
\begin{align}
&f_{nmS; nmS'}^{\mathbf{q} \mathbf{k} \mathbf{k'}} 
 = \frac{U}{N_c} \sum_{n' m' \mathbf{p}}
\frac{{m'}_{S' \mathbf{K} \downarrow}^* {m'}_{S \mathbf{K} \downarrow} 
{n'}_{S \mathbf{p} \uparrow} {n'}_{S' \mathbf{p} \uparrow}^*}
{E_{nm n'm'}^{\mathbf{q} \mathbf{k} \mathbf{k'} \mathbf{p}}},
\\
&g_{nmS; n'm'S'}^{\mathbf{q} \mathbf{k} \mathbf{k'} \mathbf{p}} 
 = -\frac{U}{N_c}
\frac{{m'}_{S' \mathbf{K} \downarrow}^* {m'}_{S \mathbf{K} \downarrow} 
{n'}_{S \mathbf{p} \uparrow} {n}_{S' \mathbf{k} \uparrow}^*}
{E_{nm n'm'}^{\mathbf{q} \mathbf{k} \mathbf{k'} \mathbf{p}}},
\\
&h_{nmS; n'm'S'}^{\mathbf{q} \mathbf{k} \mathbf{k'} \mathbf{p}} 
 = -\frac{U}{N_c}
\frac{{m'}_{S' \mathbf{K} \downarrow}^* {m'}_{S \mathbf{K} \downarrow} 
{n'}_{S \mathbf{p} \uparrow} {m}_{S' \mathbf{k'} \uparrow}^*}
{E_{nm n'm'}^{\mathbf{q} \mathbf{k} \mathbf{k'} \mathbf{p}}}.
\end{align}
Here we defined
$
\mathbf{K} = \mathbf{q} - \mathbf{k} - \mathbf{k'} - \mathbf{p}
$
and
$
E_{nm n'm'}^{\mathbf{q} \mathbf{k} \mathbf{k'} \mathbf{p}} =
\varepsilon_{n \mathbf{k} \uparrow} + \varepsilon_{m \mathbf{k'} \uparrow} 
+ \varepsilon_{n' \mathbf{p} \uparrow} + \varepsilon_{m' \mathbf{K} \downarrow} - E_4^\mathbf{q}
$
for convenience. Then we define an $N_b^3$-component vector, e.g., 
\begin{widetext}
\begin{align}
\boldsymbol{\gamma}_{\mathbf{k} \mathbf{k'}} (\mathbf{q}) = 
\begin{bmatrix}
   \gamma_{11A}^{\mathbf{k} \mathbf{k'}} (\mathbf{q}) 
& \gamma_{11B}^{\mathbf{k} \mathbf{k'}} (\mathbf{q}) 
& \gamma_{12A}^{\mathbf{k} \mathbf{k'}} (\mathbf{q})
& \gamma_{12B}^{\mathbf{k} \mathbf{k'}} (\mathbf{q}) 
& \gamma_{21A}^{\mathbf{k} \mathbf{k'}} (\mathbf{q})
& \gamma_{21B}^{\mathbf{k} \mathbf{k'}} (\mathbf{q})
& \gamma_{22A}^{\mathbf{k} \mathbf{k'}} (\mathbf{q})
& \gamma_{22B}^{\mathbf{k} \mathbf{k'}} (\mathbf{q})
& \end{bmatrix}^\mathrm{T}
\end{align}
in the case of a lattice with two sublattice sites, i.e., $N_b = 2$. Here $n = \{1, 2\}$ 
is the band index, $S = \{A, B\}$ is the sublattice index, and $\mathrm{T}$ is the transpose.
Equation~(\ref{eqn:gamma2N}) can be written as
\begin{align}
\label{eqn:gmatrix}
\boldsymbol{\gamma}_{\mathbf{k} \mathbf{k'}} (\mathbf{q}) 
= F_ \mathbf{q}^{\mathbf{k} \mathbf{k'}} 
\boldsymbol{\gamma}_{\mathbf{k} \mathbf{k'}} (\mathbf{q})
+ \sum_\mathbf{p} G_ \mathbf{q}^{\mathbf{k} \mathbf{k'} \mathbf{p}}
\boldsymbol{\gamma}_{\mathbf{p} \mathbf{k'}} (\mathbf{q})
+ \sum_\mathbf{p} H_ \mathbf{q}^{\mathbf{k} \mathbf{k'} \mathbf{p}}
\boldsymbol{\gamma}_{\mathbf{k} \mathbf{p}} (\mathbf{q})
\end{align}
where $F_ \mathbf{q}^{\mathbf{k}  \mathbf{k'}}$, 
$G_ \mathbf{q}^{\mathbf{k} \mathbf{k'} \mathbf{p}}$ 
and $H_ \mathbf{q}^{\mathbf{k} \mathbf{k'} \mathbf{p}}$ 
are $N_b^3 \times N_b^3$ matrices. For instance in the case of a lattice 
with two sublattice sites, they are given by
\begin{align}
F_\mathbf{q}^{\mathbf{k} \mathbf{k'}} &= 
\begin{pmatrix}
f_{11A;11A}^{\mathbf{q} \mathbf{k} \mathbf{k'}} &
f_{11A;11B}^{\mathbf{q} \mathbf{k} \mathbf{k'}} &
0 & 0 & 0 & 0 & 0 & 0 \\
f_{11B;11A}^{\mathbf{q} \mathbf{k} \mathbf{k'}} &
f_{11B;11B}^{\mathbf{q} \mathbf{k} \mathbf{k'}} &
0 & 0 & 0 & 0 & 0 & 0 \\
0 & 0 &
f_{12A;12A}^{\mathbf{q} \mathbf{k} \mathbf{k'}} &
f_{12A;12B}^{\mathbf{q} \mathbf{k} \mathbf{k'}} &
0 & 0 & 0 & 0  \\
0 & 0 &
f_{12B;12A}^{\mathbf{q} \mathbf{k} \mathbf{k'}} &
f_{12B;12B}^{\mathbf{q} \mathbf{k} \mathbf{k'}} &
0 & 0 & 0 & 0  \\
0 & 0 & 0 & 0 & 
f_{21A;21A}^{\mathbf{q} \mathbf{k} \mathbf{k'}} &
f_{21A;21B}^{\mathbf{q} \mathbf{k} \mathbf{k'}} &
0 & 0 \\
0 & 0 & 0 & 0 &
f_{21B;21A}^{\mathbf{q} \mathbf{k} \mathbf{k'}} &
f_{21B;21B}^{\mathbf{q} \mathbf{k} \mathbf{k'}} &
0 & 0 \\
0 & 0 & 0 & 0 & 0 & 0 &
f_{22A;22A}^{\mathbf{q} \mathbf{k} \mathbf{k'}} &
f_{22A;22B}^{\mathbf{q} \mathbf{k} \mathbf{k'}} \\
0 & 0 & 0 & 0 & 0 & 0 &
f_{22B;22A}^{\mathbf{q} \mathbf{k} \mathbf{k'}} &
f_{22B;22B}^{\mathbf{q} \mathbf{k} \mathbf{k'}}
\end{pmatrix},
\end{align}
\begin{align}
G_\mathbf{q}^{\mathbf{k} \mathbf{k'} \mathbf{p}} & = 
\begin{pmatrix}
\substack{g_{11A;11A}^{\mathbf{q} \mathbf{k} \mathbf{k'} \mathbf{p}} \\ 
\quad + g_{11A;12A}^{\mathbf{q} \mathbf{k} \mathbf{k'} \mathbf{p}}} &
\substack{g_{11A;11B}^{\mathbf{q} \mathbf{k} \mathbf{k'} \mathbf{p}} \\
\quad + g_{11A;12B}^{\mathbf{q} \mathbf{k} \mathbf{k'} \mathbf{p}}} &
0 & 0 & 
\substack{g_{11A;21A}^{\mathbf{q} \mathbf{k} \mathbf{k'} \mathbf{p}} \\
\quad + g_{11A;22A}^{\mathbf{q} \mathbf{k} \mathbf{k'} \mathbf{p}}} &
\substack{g_{11A;21B}^{\mathbf{q} \mathbf{k} \mathbf{k'} \mathbf{p}} \\
\quad + g_{11A;22B}^{\mathbf{q} \mathbf{k} \mathbf{k'} \mathbf{p}}} &
0 & 0 \\
\substack{g_{11B;11A}^{\mathbf{q} \mathbf{k} \mathbf{k'} \mathbf{p}} \\
\quad + g_{11B;12A}^{\mathbf{q} \mathbf{k} \mathbf{k'} \mathbf{p}}} &
\substack{g_{11B;11B}^{\mathbf{q} \mathbf{k} \mathbf{k'} \mathbf{p}} \\
\quad + g_{11B;12B}^{\mathbf{q} \mathbf{k} \mathbf{k'} \mathbf{p}}} &
0 & 0 & 
\substack{g_{11B;21A}^{\mathbf{q} \mathbf{k} \mathbf{k'} \mathbf{p}} \\
\quad + g_{11B;22A}^{\mathbf{q} \mathbf{k} \mathbf{k'} \mathbf{p}}} &
\substack{g_{11B;21B}^{\mathbf{q} \mathbf{k} \mathbf{k'} \mathbf{p}} \\
\quad + g_{11B;22B}^{\mathbf{q} \mathbf{k} \mathbf{k'} \mathbf{p}}} &
0 & 0 \\
0 & 0 & 
\substack{g_{12A;12A}^{\mathbf{q} \mathbf{k} \mathbf{k'} \mathbf{p}} \\ 
\quad + g_{12A;11A}^{\mathbf{q} \mathbf{k} \mathbf{k'} \mathbf{p}}} &
\substack{g_{12A;12B}^{\mathbf{q} \mathbf{k} \mathbf{k'} \mathbf{p}} \\
\quad + g_{12A;11B}^{\mathbf{q} \mathbf{k} \mathbf{k'} \mathbf{p}}} &
0 & 0 &
\substack{g_{12A;22A}^{\mathbf{q} \mathbf{k} \mathbf{k'} \mathbf{p}} \\ 
\quad + g_{12A;21A}^{\mathbf{q} \mathbf{k} \mathbf{k'} \mathbf{p}}} &
\substack{g_{12A;22B}^{\mathbf{q} \mathbf{k} \mathbf{k'} \mathbf{p}} \\
\quad + g_{12A;21B}^{\mathbf{q} \mathbf{k} \mathbf{k'} \mathbf{p}}} 
\\
0 & 0 & 
\substack{g_{12B;12A}^{\mathbf{q} \mathbf{k} \mathbf{k'} \mathbf{p}} \\ 
\quad + g_{12B;11A}^{\mathbf{q} \mathbf{k} \mathbf{k'} \mathbf{p}}} &
\substack{g_{12B;12B}^{\mathbf{q} \mathbf{k} \mathbf{k'} \mathbf{p}} \\
\quad + g_{12B;11B}^{\mathbf{q} \mathbf{k} \mathbf{k'} \mathbf{p}}} &
0 & 0 &
\substack{g_{12B;22A}^{\mathbf{q} \mathbf{k} \mathbf{k'} \mathbf{p}} \\ 
\quad + g_{12B;21A}^{\mathbf{q} \mathbf{k} \mathbf{k'} \mathbf{p}}} &
\substack{g_{12B;22B}^{\mathbf{q} \mathbf{k} \mathbf{k'} \mathbf{p}} \\
\quad + g_{12B;21B}^{\mathbf{q} \mathbf{k} \mathbf{k'} \mathbf{p}}} 
\\
\substack{g_{21A;11A}^{\mathbf{q} \mathbf{k} \mathbf{k'} \mathbf{p}} \\ 
\quad + g_{21A;12A}^{\mathbf{q} \mathbf{k} \mathbf{k'} \mathbf{p}}} &
\substack{g_{21A;11B}^{\mathbf{q} \mathbf{k} \mathbf{k'} \mathbf{p}} \\
\quad + g_{21A;12B}^{\mathbf{q} \mathbf{k} \mathbf{k'} \mathbf{p}}} &
0 & 0 & 
\substack{g_{21A;21A}^{\mathbf{q} \mathbf{k} \mathbf{k'} \mathbf{p}} \\
\quad + g_{21A;22A}^{\mathbf{q} \mathbf{k} \mathbf{k'} \mathbf{p}}} &
\substack{g_{21A;21B}^{\mathbf{q} \mathbf{k} \mathbf{k'} \mathbf{p}} \\
\quad + g_{21A;22B}^{\mathbf{q} \mathbf{k} \mathbf{k'} \mathbf{p}}} &
0 & 0 \\
\substack{g_{21B;11A}^{\mathbf{q} \mathbf{k} \mathbf{k'} \mathbf{p}} \\
\quad + g_{21B;12A}^{\mathbf{q} \mathbf{k} \mathbf{k'} \mathbf{p}}} &
\substack{g_{21B;11B}^{\mathbf{q} \mathbf{k} \mathbf{k'} \mathbf{p}} \\
\quad + g_{21B;12B}^{\mathbf{q} \mathbf{k} \mathbf{k'} \mathbf{p}}} &
0 & 0 & 
\substack{g_{21B;21A}^{\mathbf{q} \mathbf{k} \mathbf{k'} \mathbf{p}} \\
\quad + g_{21B;22A}^{\mathbf{q} \mathbf{k} \mathbf{k'} \mathbf{p}}} &
\substack{g_{21B;21B}^{\mathbf{q} \mathbf{k} \mathbf{k'} \mathbf{p}} \\
\quad + g_{21B;22B}^{\mathbf{q} \mathbf{k} \mathbf{k'} \mathbf{p}}} &
0 & 0 \\
0 & 0 & 
\substack{g_{22A;12A}^{\mathbf{q} \mathbf{k} \mathbf{k'} \mathbf{p}} \\ 
\quad + g_{22A;11A}^{\mathbf{q} \mathbf{k} \mathbf{k'} \mathbf{p}}} &
\substack{g_{22A;12B}^{\mathbf{q} \mathbf{k} \mathbf{k'} \mathbf{p}} \\
\quad + g_{22A;11B}^{\mathbf{q} \mathbf{k} \mathbf{k'} \mathbf{p}}} &
0 & 0 &
\substack{g_{22A;22A}^{\mathbf{q} \mathbf{k} \mathbf{k'} \mathbf{p}} \\ 
\quad + g_{22A;21A}^{\mathbf{q} \mathbf{k} \mathbf{k'} \mathbf{p}}} &
\substack{g_{22A;22B}^{\mathbf{q} \mathbf{k} \mathbf{k'} \mathbf{p}} \\
\quad + g_{22A;21B}^{\mathbf{q} \mathbf{k} \mathbf{k'} \mathbf{p}}} 
\\
0 & 0 & 
\substack{g_{22B;12A}^{\mathbf{q} \mathbf{k} \mathbf{k'} \mathbf{p}} \\ 
\quad + g_{22B;11A}^{\mathbf{q} \mathbf{k} \mathbf{k'} \mathbf{p}}} &
\substack{g_{22B;12B}^{\mathbf{q} \mathbf{k} \mathbf{k'} \mathbf{p}} \\
\quad + g_{22B;11B}^{\mathbf{q} \mathbf{k} \mathbf{k'} \mathbf{p}}} &
0 & 0 &
\substack{g_{22B;22A}^{\mathbf{q} \mathbf{k} \mathbf{k'} \mathbf{p}} \\ 
\quad + g_{22B;21A}^{\mathbf{q} \mathbf{k} \mathbf{k'} \mathbf{p}}} &
\substack{g_{22B;22B}^{\mathbf{q} \mathbf{k} \mathbf{k'} \mathbf{p}} \\
\quad + g_{22B;21B}^{\mathbf{q} \mathbf{k} \mathbf{k'} \mathbf{p}}} 
\end{pmatrix},
\end{align}
\begin{align}
H_\mathbf{q}^{\mathbf{k} \mathbf{k'} \mathbf{p}} & = 
\begin{pmatrix}
\substack{h_{11A;11A}^{\mathbf{q} \mathbf{k} \mathbf{k'} \mathbf{p}} \\ 
\quad + h_{11A;12A}^{\mathbf{q} \mathbf{k} \mathbf{k'} \mathbf{p}}} &
\substack{h_{11A;11B}^{\mathbf{q} \mathbf{k} \mathbf{k'} \mathbf{p}} \\
\quad + h_{11A;12B}^{\mathbf{q} \mathbf{k} \mathbf{k'} \mathbf{p}}} &
\substack{h_{11A;21A}^{\mathbf{q} \mathbf{k} \mathbf{k'} \mathbf{p}} \\ 
\quad + h_{11A;22A}^{\mathbf{q} \mathbf{k} \mathbf{k'} \mathbf{p}}} &
\substack{h_{11A;21B}^{\mathbf{q} \mathbf{k} \mathbf{k'} \mathbf{p}} \\
\quad + h_{11A;22B}^{\mathbf{q} \mathbf{k} \mathbf{k'} \mathbf{p}}} &
0 & 0 & 0 & 0 \\
\substack{h_{11B;11A}^{\mathbf{q} \mathbf{k} \mathbf{k'} \mathbf{p}} \\ 
\quad + h_{11B;12A}^{\mathbf{q} \mathbf{k} \mathbf{k'} \mathbf{p}}} &
\substack{h_{11B;11B}^{\mathbf{q} \mathbf{k} \mathbf{k'} \mathbf{p}} \\
\quad + h_{11B;12B}^{\mathbf{q} \mathbf{k} \mathbf{k'} \mathbf{p}}} &
\substack{h_{11B;21A}^{\mathbf{q} \mathbf{k} \mathbf{k'} \mathbf{p}} \\ 
\quad + h_{11B;22A}^{\mathbf{q} \mathbf{k} \mathbf{k'} \mathbf{p}}} &
\substack{h_{11B;21B}^{\mathbf{q} \mathbf{k} \mathbf{k'} \mathbf{p}} \\
\quad + h_{11B;22B}^{\mathbf{q} \mathbf{k} \mathbf{k'} \mathbf{p}}} &
0 & 0 & 0 & 0 \\
\substack{h_{12A;11A}^{\mathbf{q} \mathbf{k} \mathbf{k'} \mathbf{p}} \\ 
\quad + h_{12A;12A}^{\mathbf{q} \mathbf{k} \mathbf{k'} \mathbf{p}}} &
\substack{h_{12A;11B}^{\mathbf{q} \mathbf{k} \mathbf{k'} \mathbf{p}} \\
\quad + h_{12A;12B}^{\mathbf{q} \mathbf{k} \mathbf{k'} \mathbf{p}}} &
\substack{h_{12A;21A}^{\mathbf{q} \mathbf{k} \mathbf{k'} \mathbf{p}} \\ 
\quad + h_{12A;22A}^{\mathbf{q} \mathbf{k} \mathbf{k'} \mathbf{p}}} &
\substack{h_{12A;21B}^{\mathbf{q} \mathbf{k} \mathbf{k'} \mathbf{p}} \\
\quad + h_{12A;22B}^{\mathbf{q} \mathbf{k} \mathbf{k'} \mathbf{p}}} &
0 & 0 & 0 & 0 \\
\substack{h_{12B;11A}^{\mathbf{q} \mathbf{k} \mathbf{k'} \mathbf{p}} \\ 
\quad + h_{12B;12A}^{\mathbf{q} \mathbf{k} \mathbf{k'} \mathbf{p}}} &
\substack{h_{12B;11B}^{\mathbf{q} \mathbf{k} \mathbf{k'} \mathbf{p}} \\
\quad + h_{12B;12B}^{\mathbf{q} \mathbf{k} \mathbf{k'} \mathbf{p}}} &
\substack{h_{12B;21A}^{\mathbf{q} \mathbf{k} \mathbf{k'} \mathbf{p}} \\ 
\quad + h_{12B;22A}^{\mathbf{q} \mathbf{k} \mathbf{k'} \mathbf{p}}} &
\substack{h_{12B;21B}^{\mathbf{q} \mathbf{k} \mathbf{k'} \mathbf{p}} \\
\quad + h_{12B;22B}^{\mathbf{q} \mathbf{k} \mathbf{k'} \mathbf{p}}} &
0 & 0 & 0 & 0 \\
0 & 0 & 0 & 0 &
\substack{h_{21A;11A}^{\mathbf{q} \mathbf{k} \mathbf{k'} \mathbf{p}} \\ 
\quad + h_{21A;12A}^{\mathbf{q} \mathbf{k} \mathbf{k'} \mathbf{p}}} &
\substack{h_{21A;11B}^{\mathbf{q} \mathbf{k} \mathbf{k'} \mathbf{p}} \\
\quad + h_{21A;12B}^{\mathbf{q} \mathbf{k} \mathbf{k'} \mathbf{p}}} &
\substack{h_{21A;21A}^{\mathbf{q} \mathbf{k} \mathbf{k'} \mathbf{p}} \\ 
\quad + h_{21A;22A}^{\mathbf{q} \mathbf{k} \mathbf{k'} \mathbf{p}}} &
\substack{h_{21A;21B}^{\mathbf{q} \mathbf{k} \mathbf{k'} \mathbf{p}} \\
\quad + h_{21A;22B}^{\mathbf{q} \mathbf{k} \mathbf{k'} \mathbf{p}}} 
\\
0 & 0 & 0 & 0 &
\substack{h_{21B;11A}^{\mathbf{q} \mathbf{k} \mathbf{k'} \mathbf{p}} \\ 
\quad + h_{21B;12A}^{\mathbf{q} \mathbf{k} \mathbf{k'} \mathbf{p}}} &
\substack{h_{21B;11B}^{\mathbf{q} \mathbf{k} \mathbf{k'} \mathbf{p}} \\
\quad + h_{21B;12B}^{\mathbf{q} \mathbf{k} \mathbf{k'} \mathbf{p}}} &
\substack{h_{21B;21A}^{\mathbf{q} \mathbf{k} \mathbf{k'} \mathbf{p}} \\ 
\quad + h_{21B;22A}^{\mathbf{q} \mathbf{k} \mathbf{k'} \mathbf{p}}} &
\substack{h_{21B;21B}^{\mathbf{q} \mathbf{k} \mathbf{k'} \mathbf{p}} \\
\quad + h_{21B;22B}^{\mathbf{q} \mathbf{k} \mathbf{k'} \mathbf{p}}} 
\\
0 & 0 & 0 & 0 &
\substack{h_{22A;11A}^{\mathbf{q} \mathbf{k} \mathbf{k'} \mathbf{p}} \\ 
\quad + h_{22A;12A}^{\mathbf{q} \mathbf{k} \mathbf{k'} \mathbf{p}}} &
\substack{h_{22A;11B}^{\mathbf{q} \mathbf{k} \mathbf{k'} \mathbf{p}} \\
\quad + h_{22A;12B}^{\mathbf{q} \mathbf{k} \mathbf{k'} \mathbf{p}}} &
\substack{h_{22A;21A}^{\mathbf{q} \mathbf{k} \mathbf{k'} \mathbf{p}} \\ 
\quad + h_{22A;22A}^{\mathbf{q} \mathbf{k} \mathbf{k'} \mathbf{p}}} &
\substack{h_{22A;21B}^{\mathbf{q} \mathbf{k} \mathbf{k'} \mathbf{p}} \\
\quad + h_{22A;22B}^{\mathbf{q} \mathbf{k} \mathbf{k'} \mathbf{p}}} 
\\
0 & 0 & 0 & 0 &
\substack{h_{22B;11A}^{\mathbf{q} \mathbf{k} \mathbf{k'} \mathbf{p}} \\ 
\quad + h_{22B;12A}^{\mathbf{q} \mathbf{k} \mathbf{k'} \mathbf{p}}} &
\substack{h_{22B;11B}^{\mathbf{q} \mathbf{k} \mathbf{k'} \mathbf{p}} \\
\quad + h_{22B;12B}^{\mathbf{q} \mathbf{k} \mathbf{k'} \mathbf{p}}} &
\substack{h_{22B;21A}^{\mathbf{q} \mathbf{k} \mathbf{k'} \mathbf{p}} \\ 
\quad + h_{22B;22A}^{\mathbf{q} \mathbf{k} \mathbf{k'} \mathbf{p}}} &
\substack{h_{22B;21B}^{\mathbf{q} \mathbf{k} \mathbf{k'} \mathbf{p}} \\
\quad + h_{22B;22B}^{\mathbf{q} \mathbf{k} \mathbf{k'} \mathbf{p}}} 
\end{pmatrix}.
\end{align}
Finally we use the underlying $\mathbf{k}$-space mesh in the first BZ, i.e., 
$
\mathbf{k} = \{ \mathbf{k_1}, \mathbf{k_2}, \cdots, \mathbf{k_{N_c}} \},
$
and define an $N_b^3 N_c^2$-component vector with the following elements
\begin{align}
\Gamma_\mathbf{q} = \left[
\begin{array}{ccccccccccc}
   \boldsymbol{\gamma}_{\mathbf{k_1} \mathbf{k_1}} (\mathbf{q}) 
& \boldsymbol{\gamma}_{\mathbf{k_1} \mathbf{k_2}} (\mathbf{q})
& \cdots 
& \boldsymbol{\gamma}_{\mathbf{k_1} \mathbf{k_{N_c}}} (\mathbf{q})
& \boldsymbol{\gamma}_{\mathbf{k_2} \mathbf{k_1}} (\mathbf{q})
& \boldsymbol{\gamma}_{\mathbf{k_2} \mathbf{k_2}} (\mathbf{q}) 
& \cdots 
& \boldsymbol{\gamma}_{\mathbf{k_2} \mathbf{k_{N_c}}} (\mathbf{q})
& \boldsymbol{\gamma}_{\mathbf{k_3} \mathbf{k_1}} (\mathbf{q})
& \cdots  \cdots
& \boldsymbol{\gamma}_{\mathbf{k_{N_c}} \mathbf{k_{N_c}}} (\mathbf{q})
\end{array} \right]^\mathrm{T}.
\end{align}
Equation~(\ref{eqn:gmatrix}) can be written as
\begin{align}
\label{eqn:ggmatrix}
(\mathbb{F}_\mathbf{q} + \mathbb{G}_\mathbf{q} + \mathbb{H}_\mathbf{q}) 
\Gamma_\mathbf{q} = \Gamma_\mathbf{q},
\end{align}
where $\mathbb{F}_ \mathbf{q}$, $\mathbb{G}_ \mathbf{q}$ and $\mathbb{H}_ \mathbf{q}$
are $N_b^3 N_c^2 \times N_b^3 N_c^2$ matrices with the following elements
\begin{align}
\mathbb{F}_\mathbf{q} = \left( 
\begin{array}{cccc|cccc|cc}
F_\mathbf{q}^{\mathbf{k_1} \mathbf{k_1}} & 0 & \cdots & 0 & 
0 & 0 & \cdots & 0 & 0 & \cdots
\\
0 & F_\mathbf{q}^{\mathbf{k_1} \mathbf{k_2}} & \cdots & 0 &  
0 & 0 & \cdots & 0  & 0 & \cdots
\\
\vdots  & \vdots & \ddots & \vdots &  
\vdots  & \vdots & \ddots & \vdots & \vdots & \vdots
\\
 0 & 0 & \cdots & F_\mathbf{q}^{\mathbf{k_1} \mathbf{k_{N_c}}} & 
 0 & 0 & \cdots & 0  & 0 & \cdots
\\ \hline
0 & 0 & \cdots & 0 &
F_\mathbf{q}^{\mathbf{k_2} \mathbf{k_1}} & 0 & \cdots & 0 & 0 & \cdots
\\
0 & 0 & \cdots & 0  &
0 & F_\mathbf{q}^{\mathbf{k_2} \mathbf{k_2}} & \cdots & 0 & 0 & \cdots 
\\
\vdots  & \vdots & \ddots & \vdots &  
\vdots  & \vdots & \ddots & \vdots & \vdots & \vdots
\\
 0 & 0 & \cdots & 0  &
 0 & 0 & \cdots & F_\mathbf{q}^{\mathbf{k_2} \mathbf{k_{N_c}}} & 0 & \cdots
 \\ \hline
 0 & 0 & \cdots & 0  &
 0 & 0 & \cdots & 0 & F_\mathbf{q}^{\mathbf{k_3} \mathbf{k_1}} & \cdots
\\
 \vdots & \vdots & \cdots & \vdots  &
 \vdots & \vdots & \cdots & \vdots & \vdots & \ddots
\end{array} \right),
\end{align}
\begin{align}
\mathbb{G}_\mathbf{q} = \left(
\begin{array}{cccc|cccc|cccc|c}
   G_\mathbf{q}^{\mathbf{k_1} \mathbf{k_1} \mathbf{k_1}} & 0 & \cdots & 0 
& G_\mathbf{q}^{\mathbf{k_1} \mathbf{k_1} \mathbf{k_2}} & 0 & \cdots & 0 
& G_\mathbf{q}^{\mathbf{k_1} \mathbf{k_1} \mathbf{k_3}} & 0 & \cdots & 0 
& \cdots
\\
   0 & G_\mathbf{q}^{\mathbf{k_1} \mathbf{k_2} \mathbf{k_1}} & \cdots & 0  
& 0 & G_\mathbf{q}^{\mathbf{k_1} \mathbf{k_2} \mathbf{k_2}} & \cdots & 0 
& 0 & G_\mathbf{q}^{\mathbf{k_1} \mathbf{k_2} \mathbf{k_3}} & \cdots & 0 
& \cdots
\\
   \vdots & \vdots & \ddots & \vdots
& \vdots & \vdots & \ddots & \vdots
& \vdots & \vdots & \ddots & \vdots
& \vdots
\\
   0 & 0 & \cdots & G_\mathbf{q}^{\mathbf{k_1} \mathbf{k_{N_c}} \mathbf{k_1}} 
& 0 & 0 & \cdots & G_\mathbf{q}^{\mathbf{k_1} \mathbf{k_{N_c}} \mathbf{k_2}}
& 0 & 0 & \cdots & G_\mathbf{q}^{\mathbf{k_1} \mathbf{k_{N_c}} \mathbf{k_3}}
& \cdots
\\ \hline
   G_\mathbf{q}^{\mathbf{k_2} \mathbf{k_1} \mathbf{k_1}} & 0 & \cdots & 0 
& G_\mathbf{q}^{\mathbf{k_2} \mathbf{k_1} \mathbf{k_2}} & 0 & \cdots & 0 
& G_\mathbf{q}^{\mathbf{k_2} \mathbf{k_1} \mathbf{k_3}} & 0 & \cdots & 0 
& \cdots
\\
   0 & G_\mathbf{q}^{\mathbf{k_2} \mathbf{k_2} \mathbf{k_1}} & \cdots & 0  
& 0 & G_\mathbf{q}^{\mathbf{k_2} \mathbf{k_2} \mathbf{k_2}} & \cdots & 0 
& 0 & G_\mathbf{q}^{\mathbf{k_2} \mathbf{k_2} \mathbf{k_3}} & \cdots & 0 
& \cdots
\\
   \vdots & \vdots & \ddots & \vdots
& \vdots & \vdots & \ddots & \vdots
& \vdots & \vdots & \ddots & \vdots
& \vdots
\\
   0 & 0 & \cdots & G_\mathbf{q}^{\mathbf{k_2} \mathbf{k_{N_c}} \mathbf{k_1}} 
& 0 & 0 & \cdots & G_\mathbf{q}^{\mathbf{k_2} \mathbf{k_{N_c}} \mathbf{k_2}}
& 0 & 0 & \cdots & G_\mathbf{q}^{\mathbf{k_2} \mathbf{k_{N_c}} \mathbf{k_3}}
& \cdots
\\ \hline
   G_\mathbf{q}^{\mathbf{k_3} \mathbf{k_1} \mathbf{k_1}} & 0 & \cdots & 0 
& G_\mathbf{q}^{\mathbf{k_3} \mathbf{k_1} \mathbf{k_2}} & 0 & \cdots & 0 
& G_\mathbf{q}^{\mathbf{k_3} \mathbf{k_1} \mathbf{k_3}} & 0 & \cdots & 0 
& \cdots
\\
   0 & G_\mathbf{q}^{\mathbf{k_3} \mathbf{k_2} \mathbf{k_1}} & \cdots & 0  
& 0 & G_\mathbf{q}^{\mathbf{k_3} \mathbf{k_2} \mathbf{k_2}} & \cdots & 0 
& 0 & G_\mathbf{q}^{\mathbf{k_3} \mathbf{k_2} \mathbf{k_3}} & \cdots & 0 
& \cdots
\\
   \vdots & \vdots & \ddots & \vdots
& \vdots & \vdots & \ddots & \vdots
& \vdots & \vdots & \ddots & \vdots
& \vdots
\\
   0 & 0 & \cdots & G_\mathbf{q}^{\mathbf{k_3} \mathbf{k_{N_c}} \mathbf{k_1}} 
& 0 & 0 & \cdots & G_\mathbf{q}^{\mathbf{k_3} \mathbf{k_{N_c}} \mathbf{k_2}}
& 0 & 0 & \cdots & G_\mathbf{q}^{\mathbf{k_3} \mathbf{k_{N_c}} \mathbf{k_3}}
& \cdots
\\ \hline
   \vdots & \vdots & \cdots & \vdots
& \vdots & \vdots & \cdots & \vdots
& \vdots & \vdots & \cdots & \vdots
& \ddots
\end{array} \right),
\end{align}
\begin{align}
\mathbb{H}_\mathbf{q} = \left(
\begin{array}{cccc|cccc|ccc}
   H_\mathbf{q}^{\mathbf{k_1} \mathbf{k_1} \mathbf{k_1}} & H_\mathbf{q}^{\mathbf{k_1} \mathbf{k_1} \mathbf{k_2}} & \cdots & H_\mathbf{q}^{\mathbf{k_1} \mathbf{k_1} \mathbf{k_{N_c}}} 
& 0 & 0 & \cdots & 0 & 0 & 0 & \cdots
\\
   H_\mathbf{q}^{\mathbf{k_1} \mathbf{k_2} \mathbf{k_1}} & H_\mathbf{q}^{\mathbf{k_1} \mathbf{k_2} \mathbf{k_2}} & \cdots & H_\mathbf{q}^{\mathbf{k_1} \mathbf{k_2} \mathbf{k_{N_c}}} 
& 0 & 0 & \cdots & 0 & 0 & 0 & \cdots
\\
   \vdots & \vdots & \ddots & \vdots   
& \vdots & \vdots & \ddots & \vdots & \vdots & \vdots & \ddots
\\
   H_\mathbf{q}^{\mathbf{k_1} \mathbf{k_{N_c}} \mathbf{k_1}} & H_\mathbf{q}^{\mathbf{k_1} \mathbf{k_{N_c}} \mathbf{k_2}} & \cdots & H_\mathbf{q}^{\mathbf{k_1} \mathbf{k_{N_c}} \mathbf{k_{N_c}}}   
& 0 & 0 & \cdots & 0 & 0 & 0 & \cdots 
\\ \hline
   0 & 0 & \cdots & 0
& H_\mathbf{q}^{\mathbf{k_2} \mathbf{k_1} \mathbf{k_1}} & H_\mathbf{q}^{\mathbf{k_2} \mathbf{k_1} \mathbf{k_2}} & \cdots & H_\mathbf{q}^{\mathbf{k_2} \mathbf{k_1} \mathbf{k_{N_c}}} 
& 0 & 0 & \cdots
\\
   0 & 0 & \cdots & 0 
& H_\mathbf{q}^{\mathbf{k_2} \mathbf{k_2} \mathbf{k_1}} & H_\mathbf{q}^{\mathbf{k_2} \mathbf{k_2} \mathbf{k_2}} & \cdots & H_\mathbf{q}^{\mathbf{k_2} \mathbf{k_2} \mathbf{k_{N_c}}} 
& 0 & 0 & \cdots
\\
   \vdots & \vdots & \ddots & \vdots 
& \vdots & \vdots & \ddots & \vdots & \vdots & \vdots & \ddots 
\\
   0 & 0 & 0 & 0 
& H_\mathbf{q}^{\mathbf{k_2} \mathbf{k_{N_c}} \mathbf{k_1}} & H_\mathbf{q}^{\mathbf{k_2} \mathbf{k_{N_c}} \mathbf{k_2}} & \cdots & H_\mathbf{q}^{\mathbf{k_2} \mathbf{k_{N_c}} \mathbf{k_{N_c}}}   
& 0 & 0 & \cdots
\\ \hline
   0 & 0 & \cdots & 0 & 0 & 0 & \cdots & 0
& H_\mathbf{q}^{\mathbf{k_3} \mathbf{k_1} \mathbf{k_1}} & H_\mathbf{q}^{\mathbf{k_3} \mathbf{k_1} \mathbf{k_2}} & \cdots 
\\
   0 & 0 & \cdots & 0 & 0 & 0 & \cdots & 0
& H_\mathbf{q}^{\mathbf{k_3} \mathbf{k_2} \mathbf{k_1}} & H_\mathbf{q}^{\mathbf{k_3} \mathbf{k_2} \mathbf{k_2}} & \cdots 
\\
   \vdots & \vdots & \ddots & \vdots 
& \vdots & \vdots & \ddots & \vdots & \vdots & \vdots & \ddots
\end{array} \right).
\end{align}
\end{widetext}

Thus the four-body problem reduces to the solutions of an eigenvalue problem defined 
by Eq.~(\ref{eqn:ggmatrix}). It can be solved numerically by iterating $E_4^\mathbf{q}$ until one 
of the eigenvalues of $\mathbb{F}^\mathbf{q} + \mathbb{G}^\mathbf{q} + \mathbb{H}^\mathbf{q}$
becomes exactly $1$. Typically there are many $E_4^\mathbf{q}$ solutions for a given set 
of lattice parameters. In this work we are interested in those tetramer states with 
lowest energy for a given $\mathbf{q}$.

\section{Excited states from the exact diagonalization}
\label{sec:low}

\begin{figure*}[!htb]
    \centering
    \includegraphics[width=0.99\textwidth]{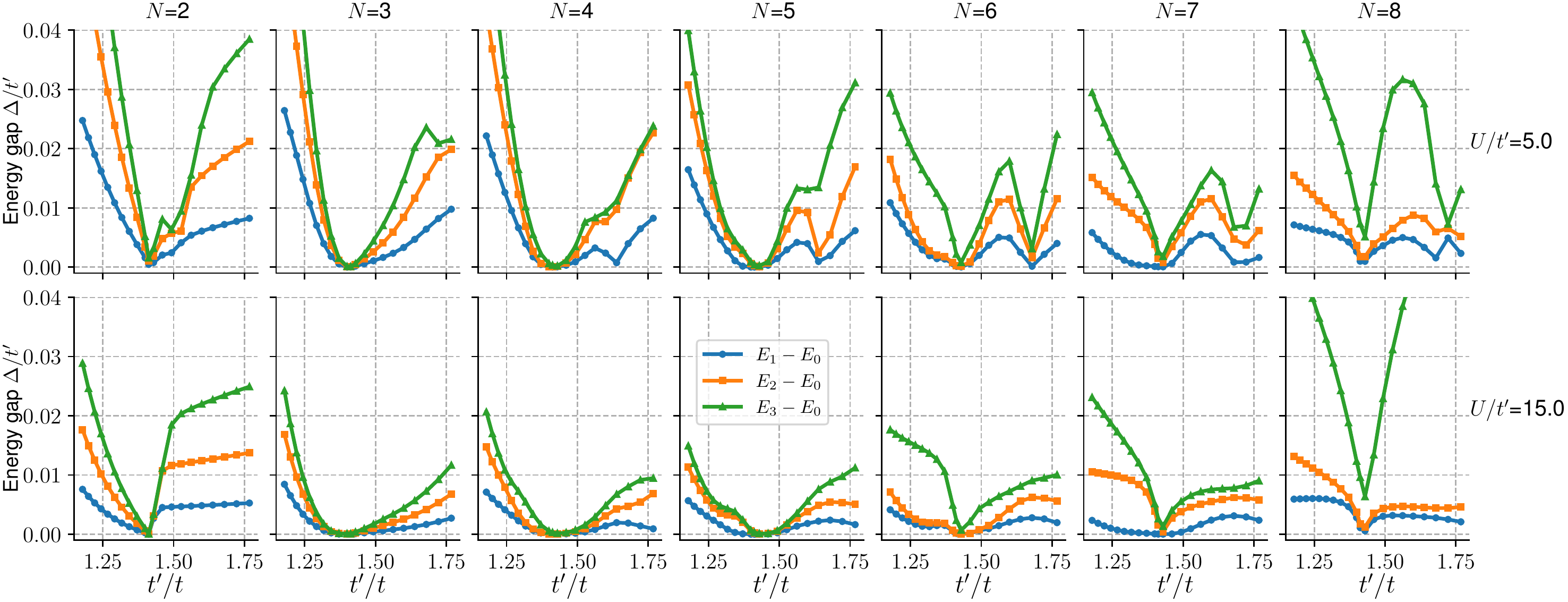}
    \caption{Energy gaps from the exact diagonalization of a lattice with $N_s=22$ sites. 
    Upper and lower panels correspond, respectively, to $U/t' = 5$ and $U/t' = 15$.
    In the flat-band case when $t'/t = \sqrt{2}$, large gaps appear once the number of
    $\uparrow$ fermions ($N$) exceeds the number of non-overlapping localized states 
    ($N_s/4$), i.e., finite-size effects appear for $N > N_s/4$.
    }
    \label{fig:ed_gaps}
\end{figure*}

As shown in Fig.~\ref{fig:ed_gaps}, the energy gaps between the first three 
excited states and the ground state vanish exactly at $t'/t = \sqrt{2}$ for 
$N = \{2, 3, 4, 5\}$. On the other hand, when the number of $\uparrow$ fermions 
$N$ exceeds the number of non-overlapping localized states in a flat band, 
one expects large energy gaps to appear due to finite-size effects. 
Here these gaps clearly appear for $N \ge 6$. In order to reveal their 
finite-size origin, we note that the localized one-body eigenstates, 
i.e., 
$
\mathcal{H}_\sigma | LS \rangle_{i \sigma} = \varepsilon_{fb} | LS \rangle_{i \sigma},
$
associated with the flat band $\varepsilon_{fb} = -2t$ in a sawtooth lattice 
can be written as~\cite{huber10, phillips15}
\begin{align}
| LS \rangle_{i \sigma} = \frac{1}{2}(\sqrt{2} c_{A i \sigma}^\dagger - c_{B i \sigma}^\dagger 
- c_{B, i-1, \sigma}^\dagger) | 0 \rangle.
\end{align}
By sketching these localized states on a lattice, one finds that there can be 
at most $N_s/4 = N_c/2$ of them that are not overlapping in space.
This is why larger and larger energy gaps appear for $N \ge 6$ in our exact 
diagonalization calculations with $N_s = 22$ sites. For this reason one may 
think of these weakly-bound $(N+1)$-body multimers as some sort of Wigner molecules 
that are caused by the occupation of these non-overlapping localized states.

In the light of these results and assuming much larger lattices, it is conceivable 
that some of these $(N+1)$-body bound states have quasi-flat dispersions when formed 
in a flat one-body band. It is important to remark here that the $(N+1)$-body bound 
states may not be energetically stable for all $N$ when $t'/t \ne \sqrt{2}$, e.g., 
see Fig.~\ref{fig:dmrg_ed}(a) for the $t'/t = \sqrt{3}$ case where $N \ge 5$ are not stable.
To understand the origin of their quasi-flat dispersions, we note that the formation 
of tightly-bound onsite trimers, onsite tetramers, etc. are all prohibited by the 
Pauli exclusion principle. For this reason the low-energy bound states of the 
$(N+1)$-body problem are necessarily offsite, 
i.e., they consist of a dimer on one site and $N-1$ monomers on other sites in the 
strong-coupling limit. These offsite trimers, offsite tetramers, etc. have negligible 
dispersions when they form in a flat band because their effective band masses are 
largely controlled by the bare effective band masses of the dimer and infinitely-massive 
monomers, as they are weakly bound. That is, the effective band mass of any offsite 
multimer is approximately given by the bare effective band mass of the dimer and
the uncoupled monomers.

\section{Fermion-boson mapping in the three-body problem}
\label{sec:fbm}

\begin{figure*}[!htb]
    \centering
    \includegraphics[width=1.98\columnwidth]{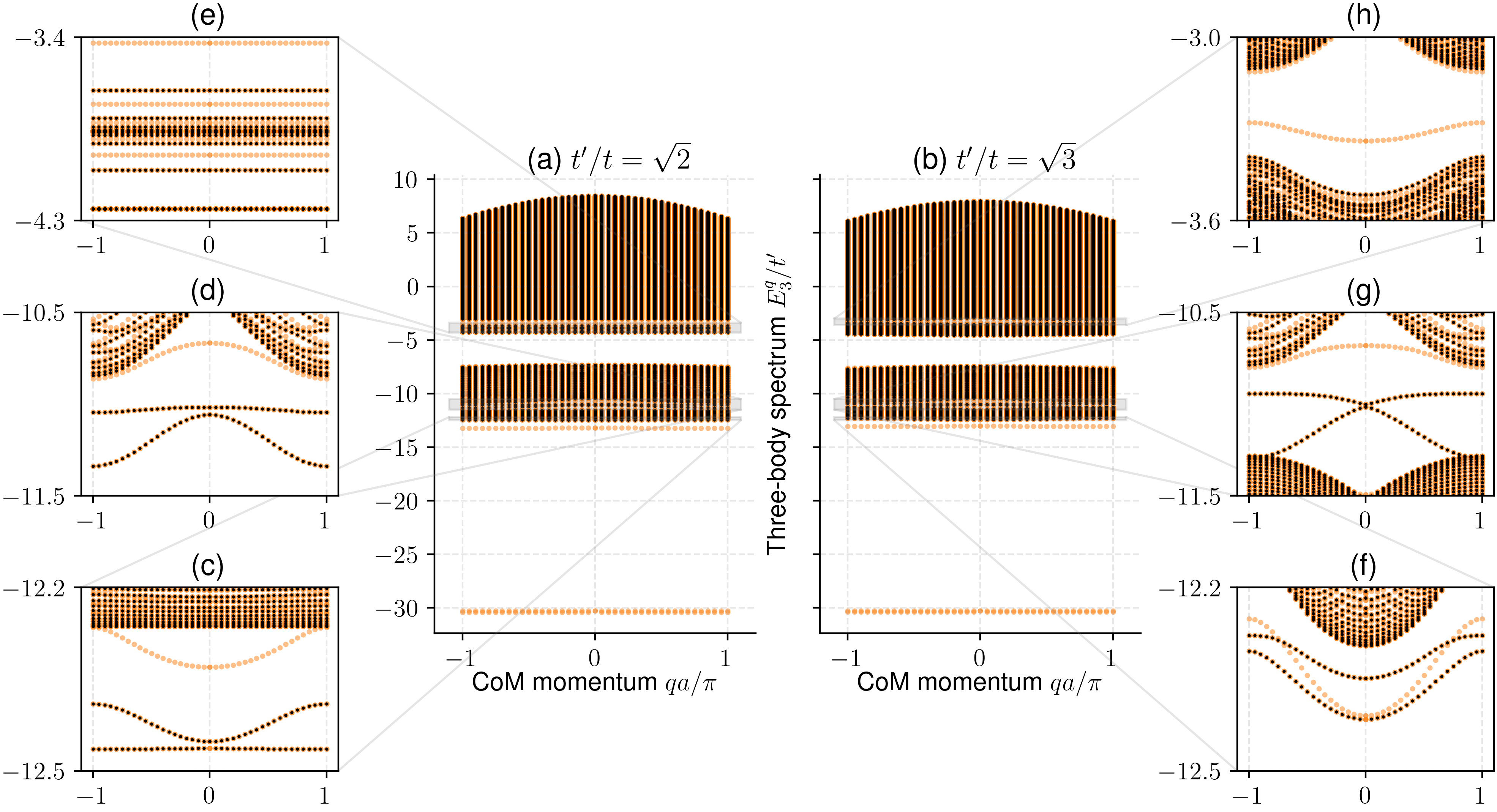} 
    \caption{Three-body spectum $E_3^q$ in a sawtooth lattice for 
    $t = t'/\sqrt{2}$ (left column) and $t = t'/\sqrt{3}$ (right column)
    when $U = 10t'$. Here $N_c = 40$ is chosen for better visibility.
    Fermion (boson) results are shown in black (yellow).
    The boson ground state is an onsite boson trimer with energy around $-3U$.
    There always exists an excited offsite boson trimer bound state for every 
    offsite fermion trimer bound state with the same energy.
    In particular insets (c) and (f) show that the ground state of the fermion 
    trimer is on top of the fifth lowest eigenvalue (i.e., third offsite boson 
    trimer branch) of the boson trimer for any given CoM momentum $q$. 
    } 
    \label{fig:figbf} 
\end{figure*}

It turns out the offsite $(2+1)$-body fermion trimers that we presented in
Sec.~\ref{sec:2p1} for the low-energy bound states of the multiband attractive 
Hubbard model are in many ways similar to the offsite boson trimers that we 
recently reported for the excited bound states of the multiband attractive 
Bose-Hubbard model~\cite{iskin22d}. 
For instance, similar to the offsite dimers, the offsite fermion and offsite 
boson trimers both have negligible dispersions when they form in a flat band. 
This is because their effective band masses are largely controlled by the 
effective band masses of the infinitely-massive monomer and the dimer,
as the two are weakly bound. 
What is striking is that, in the case of sawtooth model, the low-energy 
spectrum of the $(2+1)$-body fermion problem coincides exactly (i.e., up to the 
machine precision) with excited states of the three-boson problem. Our variational 
calculations show that this is generally the case for any given set of 
$\{t/t', U/t'\}$, and we illustrate them in Fig.~\ref{fig:figbf}.
The spectra are such that the energy of the ground fermion trimer state coincides 
with the fifth lowest eigenvalue (i.e., third offsite boson trimer branch) of 
the boson one for any given CoM momentum $q$. 
In addition the energy of the excited fermion trimer state coincides 
with the seventh lowest eigenvalue (i.e., fifth offsite boson trimer branch) 
of the boson one. 

Note that, at the bottom of the three-boson spectrum, there are also two distinct 
bound-state solutions for a given $q$~\cite{iskin22d}. They look degenerate in 
this scale with energies around $-3U$.
These are referred to as the onsite boson trimer states since their binding energy 
grows with $U$ without a limit, i.e., the three monomers are eventually tightly 
bound and they are strongly co-localized on one site in the strong-coupling limit. 
Unlike the highly dispersive onsite dimers (recall that the two-boson spectrum 
is identical to that of the two-fermion one discussed in Sec.~\ref{sec:1p1}),
the onsite trimers have nearly-flat dispersions even in the weak-binding low-$U/t'$ regime.
This is because an onsite trimer is allowed to hop in the Bose-Hubbard model 
through the so-called virtual ionization, and this brings a factor of $1/U^2$ as 
punishment from third-order perturbation theory.
Since the effective band mass of the onsite trimers is much larger in magnitude 
than that of the onsite dimers, the onsite trimers are more localized in space. 

The binding mechanism for the onsite boson trimers is very different from that 
of the offsite boson trimers. 
The binding mechanism for the strongly-bound onsite trimers (and for all 
other onsite multimers in that matter) is trivial and obvious: similar to 
the onsite dimer case, the monomers are directly bound by the onsite attraction 
term that is present in the Hamiltonian. 
On the other hand the binding mechanism for the weakly-bound offsite trimers 
(and for offsite dimers and all other offsite multimers in that matter) 
is far less obvious: the binding is mediated by a peculiar particle-exchange 
interaction between the onsite dimer and the monomer on nearest-neighbour
sites~\cite{valiente10}. 
The mediated interaction depends only on the hopping parameters $t/t'$ 
but not on $U/t'$. This is why the binding energy of the offsite boson trimers 
saturates in the strong-coupling limit.
Furthermore, given the fermion-boson mapping that is illustrated in Fig.~\ref{fig:figbf}, 
there is no doubt that the offsite fermion trimers are also bound through 
the very same mechanism. This explains why their binding energy also saturates
in the strong-coupling limit. Indeed this mechanism has recently been studied 
through a perturbation theory in the strong-coupling limit~\cite{orso22}.

\bibliography{refs}

\begin{thebibliography}{59}%
\makeatletter
\providecommand \@ifxundefined [1]{%
 \@ifx{#1\undefined}
}%
\providecommand \@ifnum [1]{%
 \ifnum #1\expandafter \@firstoftwo
 \else \expandafter \@secondoftwo
 \fi
}%
\providecommand \@ifx [1]{%
 \ifx #1\expandafter \@firstoftwo
 \else \expandafter \@secondoftwo
 \fi
}%
\providecommand \natexlab [1]{#1}%
\providecommand \enquote  [1]{``#1''}%
\providecommand \bibnamefont  [1]{#1}%
\providecommand \bibfnamefont [1]{#1}%
\providecommand \citenamefont [1]{#1}%
\providecommand \href@noop [0]{\@secondoftwo}%
\providecommand \href [0]{\begingroup \@sanitize@url \@href}%
\providecommand \@href[1]{\@@startlink{#1}\@@href}%
\providecommand \@@href[1]{\endgroup#1\@@endlink}%
\providecommand \@sanitize@url [0]{\catcode `\\12\catcode `\$12\catcode
  `\&12\catcode `\#12\catcode `\^12\catcode `\_12\catcode `\%12\relax}%
\providecommand \@@startlink[1]{}%
\providecommand \@@endlink[0]{}%
\providecommand \url  [0]{\begingroup\@sanitize@url \@url }%
\providecommand \@url [1]{\endgroup\@href {#1}{\urlprefix }}%
\providecommand \urlprefix  [0]{URL }%
\providecommand \Eprint [0]{\href }%
\providecommand \doibase [0]{https://doi.org/}%
\providecommand \selectlanguage [0]{\@gobble}%
\providecommand \bibinfo  [0]{\@secondoftwo}%
\providecommand \bibfield  [0]{\@secondoftwo}%
\providecommand \translation [1]{[#1]}%
\providecommand \BibitemOpen [0]{}%
\providecommand \bibitemStop [0]{}%
\providecommand \bibitemNoStop [0]{.\EOS\space}%
\providecommand \EOS [0]{\spacefactor3000\relax}%
\providecommand \BibitemShut  [1]{\csname bibitem#1\endcsname}%
\let\auto@bib@innerbib\@empty
\bibitem [{\citenamefont {Braaten}\ and\ \citenamefont
  {Hammer}(2006)}]{braaten06}%
  \BibitemOpen
  \bibfield  {author} {\bibinfo {author} {\bibfnamefont {E.}~\bibnamefont
  {Braaten}}\ and\ \bibinfo {author} {\bibfnamefont {H.-W.}\ \bibnamefont
  {Hammer}},\ }\bibfield  {title} {\bibinfo {title} {Universality in few-body
  systems with large scattering length},\ }\href
  {https://doi.org/10.1016/j.physrep.2006.03.001} {\bibfield  {journal}
  {\bibinfo  {journal} {Phys. Rep.}\ }\textbf {\bibinfo {volume} {428}},\
  \bibinfo {pages} {259} (\bibinfo {year} {2006})}\BibitemShut {NoStop}%
\bibitem [{\citenamefont {Hammer}\ and\ \citenamefont
  {Platter}(2010)}]{hammer10}%
  \BibitemOpen
  \bibfield  {author} {\bibinfo {author} {\bibfnamefont {H.-W.}\ \bibnamefont
  {Hammer}}\ and\ \bibinfo {author} {\bibfnamefont {L.}~\bibnamefont
  {Platter}},\ }\bibfield  {title} {\bibinfo {title} {Efimov states in nuclear
  and particle physics},\ }\href@noop {} {\bibfield  {journal} {\bibinfo
  {journal} {Annual Review of Nuclear and Particle Science}\ }\textbf {\bibinfo
  {volume} {60}},\ \bibinfo {pages} {207} (\bibinfo {year} {2010})}\BibitemShut
  {NoStop}%
\bibitem [{\citenamefont {Blume}(2012{\natexlab{a}})}]{blume12}%
  \BibitemOpen
  \bibfield  {author} {\bibinfo {author} {\bibfnamefont {D.}~\bibnamefont
  {Blume}},\ }\bibfield  {title} {\bibinfo {title} {Few-body physics with
  ultracold atomic and molecular systems in traps},\ }\href
  {https://doi.org/10.1088/0034-4885/75/4/046401} {\bibfield  {journal}
  {\bibinfo  {journal} {Rep. Prog. Phys.}\ }\textbf {\bibinfo {volume} {75}},\
  \bibinfo {pages} {046401} (\bibinfo {year} {2012}{\natexlab{a}})}\BibitemShut
  {NoStop}%
\bibitem [{\citenamefont {Greene}\ \emph {et~al.}(2017)\citenamefont {Greene},
  \citenamefont {Giannakeas},\ and\ \citenamefont {Pérez-Ríos}}]{greene17}%
  \BibitemOpen
  \bibfield  {author} {\bibinfo {author} {\bibfnamefont {C.~H.}\ \bibnamefont
  {Greene}}, \bibinfo {author} {\bibfnamefont {P.}~\bibnamefont {Giannakeas}},\
  and\ \bibinfo {author} {\bibfnamefont {J.}~\bibnamefont {Pérez-Ríos}},\
  }\bibfield  {title} {\bibinfo {title} {Universal few-body physics and cluster
  formation},\ }\href {https://doi.org/10.1103/revmodphys.89.035006} {\bibfield
   {journal} {\bibinfo  {journal} {Rev. Mod. Phys.}\ }\textbf {\bibinfo
  {volume} {89}},\ \bibinfo {pages} {035006} (\bibinfo {year}
  {2017})}\BibitemShut {NoStop}%
\bibitem [{\citenamefont {Naidon}\ and\ \citenamefont {Endo}(2017)}]{naidon17}%
  \BibitemOpen
  \bibfield  {author} {\bibinfo {author} {\bibfnamefont {P.}~\bibnamefont
  {Naidon}}\ and\ \bibinfo {author} {\bibfnamefont {S.}~\bibnamefont {Endo}},\
  }\bibfield  {title} {\bibinfo {title} {{Efimov} physics: {A} review},\ }\href
  {https://doi.org/10.1088/1361-6633/aa50e8} {\bibfield  {journal} {\bibinfo
  {journal} {Rep. Prog. Phys.}\ }\textbf {\bibinfo {volume} {80}},\ \bibinfo
  {pages} {056001} (\bibinfo {year} {2017})}\BibitemShut {NoStop}%
\bibitem [{\citenamefont {D'Incao}(2018)}]{dincao18}%
  \BibitemOpen
  \bibfield  {author} {\bibinfo {author} {\bibfnamefont {J.~P.}\ \bibnamefont
  {D'Incao}},\ }\bibfield  {title} {\bibinfo {title} {Few-body physics in
  resonantly interacting ultracold quantum gases},\ }\href
  {https://doi.org/10.1088/1361-6455/aaa116} {\bibfield  {journal} {\bibinfo
  {journal} {J. Phys. B: Atom. Mol. Opt. Phys.}\ }\textbf {\bibinfo {volume}
  {51}},\ \bibinfo {pages} {043001} (\bibinfo {year} {2018})}\BibitemShut
  {NoStop}%
\bibitem [{\citenamefont {Mattis}(1986)}]{mattis86}%
  \BibitemOpen
  \bibfield  {author} {\bibinfo {author} {\bibfnamefont {D.~C.}\ \bibnamefont
  {Mattis}},\ }\bibfield  {title} {\bibinfo {title} {The few-body problem on a
  lattice},\ }\href {https://doi.org/10.1103/revmodphys.58.361} {\bibfield
  {journal} {\bibinfo  {journal} {Rev. Mod. Phys.}\ }\textbf {\bibinfo {volume}
  {58}},\ \bibinfo {pages} {361} (\bibinfo {year} {1986})}\BibitemShut
  {NoStop}%
\bibitem [{\citenamefont {Kraemer}\ \emph {et~al.}(2006)\citenamefont
  {Kraemer}, \citenamefont {Mark}, \citenamefont {Waldburger}, \citenamefont
  {Danzl}, \citenamefont {Chin}, \citenamefont {Engeser}, \citenamefont
  {Lange}, \citenamefont {Pilch}, \citenamefont {Jaakkola}, \citenamefont
  {Nägerl},\ and\ \citenamefont {Grimm}}]{kraemer06}%
  \BibitemOpen
  \bibfield  {author} {\bibinfo {author} {\bibfnamefont {T.}~\bibnamefont
  {Kraemer}}, \bibinfo {author} {\bibfnamefont {M.}~\bibnamefont {Mark}},
  \bibinfo {author} {\bibfnamefont {P.}~\bibnamefont {Waldburger}}, \bibinfo
  {author} {\bibfnamefont {J.~G.}\ \bibnamefont {Danzl}}, \bibinfo {author}
  {\bibfnamefont {C.}~\bibnamefont {Chin}}, \bibinfo {author} {\bibfnamefont
  {B.}~\bibnamefont {Engeser}}, \bibinfo {author} {\bibfnamefont {A.~D.}\
  \bibnamefont {Lange}}, \bibinfo {author} {\bibfnamefont {K.}~\bibnamefont
  {Pilch}}, \bibinfo {author} {\bibfnamefont {A.}~\bibnamefont {Jaakkola}},
  \bibinfo {author} {\bibfnamefont {H.-C.}\ \bibnamefont {Nägerl}},\ and\
  \bibinfo {author} {\bibfnamefont {R.}~\bibnamefont {Grimm}},\ }\bibfield
  {title} {\bibinfo {title} {Evidence for {Efimov} quantum states in an
  ultracold gas of caesium atoms},\ }\href
  {https://doi.org/10.1038/nature04626} {\bibfield  {journal} {\bibinfo
  {journal} {Nature}\ }\textbf {\bibinfo {volume} {440}},\ \bibinfo {pages}
  {315} (\bibinfo {year} {2006})}\BibitemShut {NoStop}%
\bibitem [{\citenamefont {Zaccanti}\ \emph {et~al.}(2009)\citenamefont
  {Zaccanti}, \citenamefont {Deissler}, \citenamefont {D’Errico},
  \citenamefont {Fattori}, \citenamefont {Jona-Lasinio}, \citenamefont
  {M{\"u}ller}, \citenamefont {Roati}, \citenamefont {Inguscio},\ and\
  \citenamefont {Modugno}}]{zaccanti09}%
  \BibitemOpen
  \bibfield  {author} {\bibinfo {author} {\bibfnamefont {M.}~\bibnamefont
  {Zaccanti}}, \bibinfo {author} {\bibfnamefont {B.}~\bibnamefont {Deissler}},
  \bibinfo {author} {\bibfnamefont {C.}~\bibnamefont {D’Errico}}, \bibinfo
  {author} {\bibfnamefont {M.}~\bibnamefont {Fattori}}, \bibinfo {author}
  {\bibfnamefont {M.}~\bibnamefont {Jona-Lasinio}}, \bibinfo {author}
  {\bibfnamefont {S.}~\bibnamefont {M{\"u}ller}}, \bibinfo {author}
  {\bibfnamefont {G.}~\bibnamefont {Roati}}, \bibinfo {author} {\bibfnamefont
  {M.}~\bibnamefont {Inguscio}},\ and\ \bibinfo {author} {\bibfnamefont
  {G.}~\bibnamefont {Modugno}},\ }\bibfield  {title} {\bibinfo {title}
  {Observation of an {Efimov} spectrum in an atomic system},\ }\href
  {https://doi.org/10.1038/nphys1334} {\bibfield  {journal} {\bibinfo
  {journal} {Nat. Phys.}\ }\textbf {\bibinfo {volume} {5}},\ \bibinfo {pages}
  {586} (\bibinfo {year} {2009})}\BibitemShut {NoStop}%
\bibitem [{\citenamefont {Pollack}\ \emph {et~al.}(2009)\citenamefont
  {Pollack}, \citenamefont {Dries},\ and\ \citenamefont {Hulet}}]{pollack09}%
  \BibitemOpen
  \bibfield  {author} {\bibinfo {author} {\bibfnamefont {S.~E.}\ \bibnamefont
  {Pollack}}, \bibinfo {author} {\bibfnamefont {D.}~\bibnamefont {Dries}},\
  and\ \bibinfo {author} {\bibfnamefont {R.~G.}\ \bibnamefont {Hulet}},\
  }\bibfield  {title} {\bibinfo {title} {Universality in three- and four-body
  bound states of ultracold atoms},\ }\href
  {https://doi.org/10.1126/science.1182840} {\bibfield  {journal} {\bibinfo
  {journal} {Science}\ }\textbf {\bibinfo {volume} {326}},\ \bibinfo {pages}
  {1683} (\bibinfo {year} {2009})}\BibitemShut {NoStop}%
\bibitem [{\citenamefont {Gross}\ \emph {et~al.}(2009)\citenamefont {Gross},
  \citenamefont {Shotan}, \citenamefont {Kokkelmans},\ and\ \citenamefont
  {Khaykovich}}]{gross09}%
  \BibitemOpen
  \bibfield  {author} {\bibinfo {author} {\bibfnamefont {N.}~\bibnamefont
  {Gross}}, \bibinfo {author} {\bibfnamefont {Z.}~\bibnamefont {Shotan}},
  \bibinfo {author} {\bibfnamefont {S.}~\bibnamefont {Kokkelmans}},\ and\
  \bibinfo {author} {\bibfnamefont {L.}~\bibnamefont {Khaykovich}},\ }\bibfield
   {title} {\bibinfo {title} {Observation of universality in ultracold
  $^{7}\mathrm{Li}$ three-body recombination},\ }\href
  {https://doi.org/10.1103/PhysRevLett.103.163202} {\bibfield  {journal}
  {\bibinfo  {journal} {Phys. Rev. Lett.}\ }\textbf {\bibinfo {volume} {103}},\
  \bibinfo {pages} {163202} (\bibinfo {year} {2009})}\BibitemShut {NoStop}%
\bibitem [{\citenamefont {Grimm}(2019)}]{grimm19}%
  \BibitemOpen
  \bibfield  {author} {\bibinfo {author} {\bibfnamefont {R.}~\bibnamefont
  {Grimm}},\ }\bibfield  {title} {\bibinfo {title} {{Efimov} states in an
  ultracold gas: How it happened in the laboratory},\ }\href@noop {} {\bibfield
   {journal} {\bibinfo  {journal} {Few-Body Systems}\ }\textbf {\bibinfo
  {volume} {60}},\ \bibinfo {pages} {1} (\bibinfo {year} {2019})}\BibitemShut
  {NoStop}%
\bibitem [{\citenamefont {Kartavtsev}\ and\ \citenamefont
  {Malykh}(2007)}]{kartavtsev07}%
  \BibitemOpen
  \bibfield  {author} {\bibinfo {author} {\bibfnamefont {O.~I.}\ \bibnamefont
  {Kartavtsev}}\ and\ \bibinfo {author} {\bibfnamefont {A.~V.}\ \bibnamefont
  {Malykh}},\ }\bibfield  {title} {\bibinfo {title} {Low-energy three-body
  dynamics in binary quantum gases},\ }\href
  {https://doi.org/10.1088/0953-4075/40/7/011} {\bibfield  {journal} {\bibinfo
  {journal} {J. Phys. B: Atom. Mol. Opt. Phys.}\ }\textbf {\bibinfo {volume}
  {40}},\ \bibinfo {pages} {1429} (\bibinfo {year} {2007})}\BibitemShut
  {NoStop}%
\bibitem [{\citenamefont {Castin}\ \emph {et~al.}(2010)\citenamefont {Castin},
  \citenamefont {Mora},\ and\ \citenamefont {Pricoupenko}}]{castin10}%
  \BibitemOpen
  \bibfield  {author} {\bibinfo {author} {\bibfnamefont {Y.}~\bibnamefont
  {Castin}}, \bibinfo {author} {\bibfnamefont {C.}~\bibnamefont {Mora}},\ and\
  \bibinfo {author} {\bibfnamefont {L.}~\bibnamefont {Pricoupenko}},\
  }\bibfield  {title} {\bibinfo {title} {Four-body {Efimov} effect for three
  fermions and a lighter particle},\ }\href
  {https://doi.org/10.1103/PhysRevLett.105.223201} {\bibfield  {journal}
  {\bibinfo  {journal} {Phys. Rev. Lett.}\ }\textbf {\bibinfo {volume} {105}},\
  \bibinfo {pages} {223201} (\bibinfo {year} {2010})}\BibitemShut {NoStop}%
\bibitem [{\citenamefont {Levinsen}\ and\ \citenamefont
  {Parish}(2013)}]{levinsen13}%
  \BibitemOpen
  \bibfield  {author} {\bibinfo {author} {\bibfnamefont {J.}~\bibnamefont
  {Levinsen}}\ and\ \bibinfo {author} {\bibfnamefont {M.~M.}\ \bibnamefont
  {Parish}},\ }\bibfield  {title} {\bibinfo {title} {Bound states in a
  quasi-two-dimensional {Fermi} gas},\ }\href
  {https://doi.org/10.1103/PhysRevLett.110.055304} {\bibfield  {journal}
  {\bibinfo  {journal} {Phys. Rev. Lett.}\ }\textbf {\bibinfo {volume} {110}},\
  \bibinfo {pages} {055304} (\bibinfo {year} {2013})}\BibitemShut {NoStop}%
\bibitem [{\citenamefont {Jag}\ \emph {et~al.}(2014)\citenamefont {Jag},
  \citenamefont {Zaccanti}, \citenamefont {Cetina}, \citenamefont {Lous},
  \citenamefont {Schreck}, \citenamefont {Grimm}, \citenamefont {Petrov},\ and\
  \citenamefont {Levinsen}}]{jag14}%
  \BibitemOpen
  \bibfield  {author} {\bibinfo {author} {\bibfnamefont {M.}~\bibnamefont
  {Jag}}, \bibinfo {author} {\bibfnamefont {M.}~\bibnamefont {Zaccanti}},
  \bibinfo {author} {\bibfnamefont {M.}~\bibnamefont {Cetina}}, \bibinfo
  {author} {\bibfnamefont {R.~S.}\ \bibnamefont {Lous}}, \bibinfo {author}
  {\bibfnamefont {F.}~\bibnamefont {Schreck}}, \bibinfo {author} {\bibfnamefont
  {R.}~\bibnamefont {Grimm}}, \bibinfo {author} {\bibfnamefont {D.~S.}\
  \bibnamefont {Petrov}},\ and\ \bibinfo {author} {\bibfnamefont
  {J.}~\bibnamefont {Levinsen}},\ }\bibfield  {title} {\bibinfo {title}
  {Observation of a strong atom-dimer attraction in a mass-imbalanced
  {Fermi}-{Fermi} mixture},\ }\href
  {https://doi.org/10.1103/PhysRevLett.112.075302} {\bibfield  {journal}
  {\bibinfo  {journal} {Phys. Rev. Lett.}\ }\textbf {\bibinfo {volume} {112}},\
  \bibinfo {pages} {075302} (\bibinfo {year} {2014})}\BibitemShut {NoStop}%
\bibitem [{\citenamefont {Blume}(2012{\natexlab{b}})}]{blume12b}%
  \BibitemOpen
  \bibfield  {author} {\bibinfo {author} {\bibfnamefont {D.}~\bibnamefont
  {Blume}},\ }\bibfield  {title} {\bibinfo {title} {Universal four-body states
  in heavy-light mixtures with a positive scattering length},\ }\href
  {https://doi.org/10.1103/PhysRevLett.109.230404} {\bibfield  {journal}
  {\bibinfo  {journal} {Phys. Rev. Lett.}\ }\textbf {\bibinfo {volume} {109}},\
  \bibinfo {pages} {230404} (\bibinfo {year} {2012}{\natexlab{b}})}\BibitemShut
  {NoStop}%
\bibitem [{\citenamefont {Bazak}\ and\ \citenamefont {Petrov}(2017)}]{bazak17}%
  \BibitemOpen
  \bibfield  {author} {\bibinfo {author} {\bibfnamefont {B.}~\bibnamefont
  {Bazak}}\ and\ \bibinfo {author} {\bibfnamefont {D.~S.}\ \bibnamefont
  {Petrov}},\ }\bibfield  {title} {\bibinfo {title} {Five-body {Efimov} effect
  and universal pentamer in fermionic mixtures},\ }\href
  {https://doi.org/10.1103/PhysRevLett.118.083002} {\bibfield  {journal}
  {\bibinfo  {journal} {Phys. Rev. Lett.}\ }\textbf {\bibinfo {volume} {118}},\
  \bibinfo {pages} {083002} (\bibinfo {year} {2017})}\BibitemShut {NoStop}%
\bibitem [{\citenamefont {Sanayei}\ \emph {et~al.}(2020)\citenamefont
  {Sanayei}, \citenamefont {Naidon},\ and\ \citenamefont {Mathey}}]{sanayei20}%
  \BibitemOpen
  \bibfield  {author} {\bibinfo {author} {\bibfnamefont {A.}~\bibnamefont
  {Sanayei}}, \bibinfo {author} {\bibfnamefont {P.}~\bibnamefont {Naidon}},\
  and\ \bibinfo {author} {\bibfnamefont {L.}~\bibnamefont {Mathey}},\
  }\bibfield  {title} {\bibinfo {title} {Electron trimer states in conventional
  superconductors},\ }\href {https://doi.org/10.1103/PhysRevResearch.2.013341}
  {\bibfield  {journal} {\bibinfo  {journal} {Phys. Rev. Research}\ }\textbf
  {\bibinfo {volume} {2}},\ \bibinfo {pages} {013341} (\bibinfo {year}
  {2020})}\BibitemShut {NoStop}%
\bibitem [{\citenamefont {Liu}\ \emph {et~al.}(2022)\citenamefont {Liu},
  \citenamefont {Peng},\ and\ \citenamefont {Cui}}]{liu22}%
  \BibitemOpen
  \bibfield  {author} {\bibinfo {author} {\bibfnamefont {R.}~\bibnamefont
  {Liu}}, \bibinfo {author} {\bibfnamefont {C.}~\bibnamefont {Peng}},\ and\
  \bibinfo {author} {\bibfnamefont {X.}~\bibnamefont {Cui}},\ }\href@noop {}
  {\bibinfo {title} {Universal tetramer and pentamer in two-dimensional
  fermionic mixtures}} (\bibinfo {year} {2022}),\ \Eprint
  {https://arxiv.org/abs/2202.01437} {arXiv:2202.01437} \BibitemShut {NoStop}%
\bibitem [{\citenamefont {Orso}\ \emph {et~al.}(2010)\citenamefont {Orso},
  \citenamefont {Burovski},\ and\ \citenamefont {Jolicoeur}}]{orso10}%
  \BibitemOpen
  \bibfield  {author} {\bibinfo {author} {\bibfnamefont {G.}~\bibnamefont
  {Orso}}, \bibinfo {author} {\bibfnamefont {E.}~\bibnamefont {Burovski}},\
  and\ \bibinfo {author} {\bibfnamefont {T.}~\bibnamefont {Jolicoeur}},\
  }\bibfield  {title} {\bibinfo {title} {Luttinger liquid of trimers in {Fermi}
  gases with unequal masses},\ }\href
  {https://doi.org/10.1103/PhysRevLett.104.065301} {\bibfield  {journal}
  {\bibinfo  {journal} {Phys. Rev. Lett.}\ }\textbf {\bibinfo {volume} {104}},\
  \bibinfo {pages} {065301} (\bibinfo {year} {2010})}\BibitemShut {NoStop}%
\bibitem [{\citenamefont {Orso}\ \emph {et~al.}(2011)\citenamefont {Orso},
  \citenamefont {Burovski},\ and\ \citenamefont {Jolicoeur}}]{orso11}%
  \BibitemOpen
  \bibfield  {author} {\bibinfo {author} {\bibfnamefont {G.}~\bibnamefont
  {Orso}}, \bibinfo {author} {\bibfnamefont {E.}~\bibnamefont {Burovski}},\
  and\ \bibinfo {author} {\bibfnamefont {T.}~\bibnamefont {Jolicoeur}},\
  }\bibfield  {title} {\bibinfo {title} {Fermionic trimers in spin-dependent
  optical lattices},\ }\href {https://doi.org/10.1016/j.crhy.2010.10.008}
  {\bibfield  {journal} {\bibinfo  {journal} {Cr. Phys.}\ }\textbf {\bibinfo
  {volume} {12}},\ \bibinfo {pages} {39} (\bibinfo {year} {2011})}\BibitemShut
  {NoStop}%
\bibitem [{\citenamefont {Roux}\ \emph {et~al.}(2011)\citenamefont {Roux},
  \citenamefont {Burovski},\ and\ \citenamefont {Jolicoeur}}]{roux11}%
  \BibitemOpen
  \bibfield  {author} {\bibinfo {author} {\bibfnamefont {G.}~\bibnamefont
  {Roux}}, \bibinfo {author} {\bibfnamefont {E.}~\bibnamefont {Burovski}},\
  and\ \bibinfo {author} {\bibfnamefont {T.}~\bibnamefont {Jolicoeur}},\
  }\bibfield  {title} {\bibinfo {title} {Multimer formation in one-dimensional
  two-component gases and trimer phase in the asymmetric attractive {Hubbard}
  model},\ }\href {https://doi.org/10.1103/PhysRevA.83.053618} {\bibfield
  {journal} {\bibinfo  {journal} {Phys. Rev. A}\ }\textbf {\bibinfo {volume}
  {83}},\ \bibinfo {pages} {053618} (\bibinfo {year} {2011})}\BibitemShut
  {NoStop}%
\bibitem [{\citenamefont {Dalmonte}\ \emph {et~al.}(2012)\citenamefont
  {Dalmonte}, \citenamefont {Dieckmann}, \citenamefont {Roscilde},
  \citenamefont {Hartl}, \citenamefont {Feiguin}, \citenamefont
  {Schollw\"ock},\ and\ \citenamefont {Heidrich-Meisner}}]{dalmonte12}%
  \BibitemOpen
  \bibfield  {author} {\bibinfo {author} {\bibfnamefont {M.}~\bibnamefont
  {Dalmonte}}, \bibinfo {author} {\bibfnamefont {K.}~\bibnamefont {Dieckmann}},
  \bibinfo {author} {\bibfnamefont {T.}~\bibnamefont {Roscilde}}, \bibinfo
  {author} {\bibfnamefont {C.}~\bibnamefont {Hartl}}, \bibinfo {author}
  {\bibfnamefont {A.~E.}\ \bibnamefont {Feiguin}}, \bibinfo {author}
  {\bibfnamefont {U.}~\bibnamefont {Schollw\"ock}},\ and\ \bibinfo {author}
  {\bibfnamefont {F.}~\bibnamefont {Heidrich-Meisner}},\ }\bibfield  {title}
  {\bibinfo {title} {Dimer, trimer, and {Fulde-Ferrell-Larkin-Ovchinnikov}
  liquids in mass- and spin-imbalanced trapped binary mixtures in one
  dimension},\ }\href {https://doi.org/10.1103/PhysRevA.85.063608} {\bibfield
  {journal} {\bibinfo  {journal} {Phys. Rev. A}\ }\textbf {\bibinfo {volume}
  {85}},\ \bibinfo {pages} {063608} (\bibinfo {year} {2012})}\BibitemShut
  {NoStop}%
\bibitem [{\citenamefont {Dhar}\ \emph {et~al.}(2018)\citenamefont {Dhar},
  \citenamefont {T\"orm\"a},\ and\ \citenamefont {Kinnunen}}]{dhar18}%
  \BibitemOpen
  \bibfield  {author} {\bibinfo {author} {\bibfnamefont {A.}~\bibnamefont
  {Dhar}}, \bibinfo {author} {\bibfnamefont {P.}~\bibnamefont {T\"orm\"a}},\
  and\ \bibinfo {author} {\bibfnamefont {J.~J.}\ \bibnamefont {Kinnunen}},\
  }\bibfield  {title} {\bibinfo {title} {Fast trimers in a one-dimensional
  extended {Fermi}-{Hubbard} model},\ }\href
  {https://doi.org/10.1103/PhysRevA.97.043624} {\bibfield  {journal} {\bibinfo
  {journal} {Phys. Rev. A}\ }\textbf {\bibinfo {volume} {97}},\ \bibinfo
  {pages} {043624} (\bibinfo {year} {2018})}\BibitemShut {NoStop}%
\bibitem [{\citenamefont {Takahashi}(1970)}]{takahashi70}%
  \BibitemOpen
  \bibfield  {author} {\bibinfo {author} {\bibfnamefont {M.}~\bibnamefont
  {Takahashi}},\ }\bibfield  {title} {\bibinfo {title} {{Excitonic Insulator in
  One Dimension}},\ }\href {https://doi.org/10.1143/PTP.43.917} {\bibfield
  {journal} {\bibinfo  {journal} {Prog. Theor. Phys.}\ }\textbf {\bibinfo
  {volume} {43}},\ \bibinfo {pages} {917} (\bibinfo {year} {1970})}\BibitemShut
  {NoStop}%
\bibitem [{\citenamefont {Orso}\ and\ \citenamefont {Singh}(2022)}]{orso22}%
  \BibitemOpen
  \bibfield  {author} {\bibinfo {author} {\bibfnamefont {G.}~\bibnamefont
  {Orso}}\ and\ \bibinfo {author} {\bibfnamefont {M.}~\bibnamefont {Singh}},\
  }\bibfield  {title} {\bibinfo {title} {Pairs, trimers, and bcs-bec crossover
  near a flat band: Sawtooth lattice},\ }\href
  {https://doi.org/10.1103/PhysRevB.106.014504} {\bibfield  {journal} {\bibinfo
   {journal} {Phys. Rev. B}\ }\textbf {\bibinfo {volume} {106}},\ \bibinfo
  {pages} {014504} (\bibinfo {year} {2022})}\BibitemShut {NoStop}%
\bibitem [{\citenamefont {Iskin}(2022{\natexlab{a}})}]{iskin22b}%
  \BibitemOpen
  \bibfield  {author} {\bibinfo {author} {\bibfnamefont {M.}~\bibnamefont
  {Iskin}},\ }\bibfield  {title} {\bibinfo {title} {Three-body problem in a
  multiband hubbard model},\ }\href
  {https://doi.org/10.1103/PhysRevA.105.063310} {\bibfield  {journal} {\bibinfo
   {journal} {Phys. Rev. A}\ }\textbf {\bibinfo {volume} {105}},\ \bibinfo
  {pages} {063310} (\bibinfo {year} {2022}{\natexlab{a}})}\BibitemShut
  {NoStop}%
\bibitem [{\citenamefont {Jo}\ \emph {et~al.}(2012)\citenamefont {Jo},
  \citenamefont {Guzman}, \citenamefont {Thomas}, \citenamefont {Hosur},
  \citenamefont {Vishwanath},\ and\ \citenamefont {Stamper-Kurn}}]{jo12}%
  \BibitemOpen
  \bibfield  {author} {\bibinfo {author} {\bibfnamefont {G.-B.}\ \bibnamefont
  {Jo}}, \bibinfo {author} {\bibfnamefont {J.}~\bibnamefont {Guzman}}, \bibinfo
  {author} {\bibfnamefont {C.~K.}\ \bibnamefont {Thomas}}, \bibinfo {author}
  {\bibfnamefont {P.}~\bibnamefont {Hosur}}, \bibinfo {author} {\bibfnamefont
  {A.}~\bibnamefont {Vishwanath}},\ and\ \bibinfo {author} {\bibfnamefont
  {D.~M.}\ \bibnamefont {Stamper-Kurn}},\ }\bibfield  {title} {\bibinfo {title}
  {Ultracold atoms in a tunable optical kagome lattice},\ }\href
  {https://doi.org/10.1103/physrevlett.108.045305} {\bibfield  {journal}
  {\bibinfo  {journal} {Phys. Rev. Lett.}\ }\textbf {\bibinfo {volume} {108}},\
  \bibinfo {pages} {045305} (\bibinfo {year} {2012})}\BibitemShut {NoStop}%
\bibitem [{\citenamefont {Nakata}\ \emph {et~al.}(2012)\citenamefont {Nakata},
  \citenamefont {Okada}, \citenamefont {Nakanishi},\ and\ \citenamefont
  {Kitano}}]{nakata12}%
  \BibitemOpen
  \bibfield  {author} {\bibinfo {author} {\bibfnamefont {Y.}~\bibnamefont
  {Nakata}}, \bibinfo {author} {\bibfnamefont {T.}~\bibnamefont {Okada}},
  \bibinfo {author} {\bibfnamefont {T.}~\bibnamefont {Nakanishi}},\ and\
  \bibinfo {author} {\bibfnamefont {M.}~\bibnamefont {Kitano}},\ }\bibfield
  {title} {\bibinfo {title} {Observation of flat band for terahertz spoof
  plasmons in a metallic kagomé lattice},\ }\href
  {https://doi.org/10.1103/physrevb.85.205128} {\bibfield  {journal} {\bibinfo
  {journal} {Phys. Rev. B}\ }\textbf {\bibinfo {volume} {85}},\ \bibinfo
  {pages} {205128} (\bibinfo {year} {2012})}\BibitemShut {NoStop}%
\bibitem [{\citenamefont {Li}\ \emph {et~al.}(2018)\citenamefont {Li},
  \citenamefont {Zhuang}, \citenamefont {Wang}, \citenamefont {Feng},
  \citenamefont {Gao}, \citenamefont {Xu}, \citenamefont {Hao}, \citenamefont
  {Wang}, \citenamefont {Zhang}, \citenamefont {Wu}, \citenamefont {Dou},
  \citenamefont {Chen}, \citenamefont {Hu},\ and\ \citenamefont {Du}}]{li18}%
  \BibitemOpen
  \bibfield  {author} {\bibinfo {author} {\bibfnamefont {Z.}~\bibnamefont
  {Li}}, \bibinfo {author} {\bibfnamefont {J.}~\bibnamefont {Zhuang}}, \bibinfo
  {author} {\bibfnamefont {L.}~\bibnamefont {Wang}}, \bibinfo {author}
  {\bibfnamefont {H.}~\bibnamefont {Feng}}, \bibinfo {author} {\bibfnamefont
  {Q.}~\bibnamefont {Gao}}, \bibinfo {author} {\bibfnamefont {X.}~\bibnamefont
  {Xu}}, \bibinfo {author} {\bibfnamefont {W.}~\bibnamefont {Hao}}, \bibinfo
  {author} {\bibfnamefont {X.}~\bibnamefont {Wang}}, \bibinfo {author}
  {\bibfnamefont {C.}~\bibnamefont {Zhang}}, \bibinfo {author} {\bibfnamefont
  {K.}~\bibnamefont {Wu}}, \bibinfo {author} {\bibfnamefont {S.~X.}\
  \bibnamefont {Dou}}, \bibinfo {author} {\bibfnamefont {L.}~\bibnamefont
  {Chen}}, \bibinfo {author} {\bibfnamefont {Z.}~\bibnamefont {Hu}},\ and\
  \bibinfo {author} {\bibfnamefont {Y.}~\bibnamefont {Du}},\ }\bibfield
  {title} {\bibinfo {title} {Realization of flat band with possible nontrivial
  topology in electronic kagome lattice},\ }\href
  {https://doi.org/10.1126/sciadv.aau4511} {\bibfield  {journal} {\bibinfo
  {journal} {Science Advances}\ }\textbf {\bibinfo {volume} {4}} (\bibinfo
  {year} {2018})}\BibitemShut {NoStop}%
\bibitem [{\citenamefont {Diebel}\ \emph {et~al.}(2016)\citenamefont {Diebel},
  \citenamefont {Leykam}, \citenamefont {Kroesen}, \citenamefont {Denz},\ and\
  \citenamefont {Desyatnikov}}]{diebel16}%
  \BibitemOpen
  \bibfield  {author} {\bibinfo {author} {\bibfnamefont {F.}~\bibnamefont
  {Diebel}}, \bibinfo {author} {\bibfnamefont {D.}~\bibnamefont {Leykam}},
  \bibinfo {author} {\bibfnamefont {S.}~\bibnamefont {Kroesen}}, \bibinfo
  {author} {\bibfnamefont {C.}~\bibnamefont {Denz}},\ and\ \bibinfo {author}
  {\bibfnamefont {A.~S.}\ \bibnamefont {Desyatnikov}},\ }\bibfield  {title}
  {\bibinfo {title} {Conical diffraction and composite {Lieb} bosons in
  photonic lattices},\ }\href {https://doi.org/10.1103/physrevlett.116.183902}
  {\bibfield  {journal} {\bibinfo  {journal} {Phys. Rev. Lett.}\ }\textbf
  {\bibinfo {volume} {116}},\ \bibinfo {pages} {183902} (\bibinfo {year}
  {2016})}\BibitemShut {NoStop}%
\bibitem [{\citenamefont {Kajiwara}\ \emph {et~al.}(2016)\citenamefont
  {Kajiwara}, \citenamefont {Urade}, \citenamefont {Nakata}, \citenamefont
  {Nakanishi},\ and\ \citenamefont {Kitano}}]{kajiwara16}%
  \BibitemOpen
  \bibfield  {author} {\bibinfo {author} {\bibfnamefont {S.}~\bibnamefont
  {Kajiwara}}, \bibinfo {author} {\bibfnamefont {Y.}~\bibnamefont {Urade}},
  \bibinfo {author} {\bibfnamefont {Y.}~\bibnamefont {Nakata}}, \bibinfo
  {author} {\bibfnamefont {T.}~\bibnamefont {Nakanishi}},\ and\ \bibinfo
  {author} {\bibfnamefont {M.}~\bibnamefont {Kitano}},\ }\bibfield  {title}
  {\bibinfo {title} {Observation of a nonradiative flat band for spoof surface
  plasmons in a metallic {Lieb} lattice},\ }\href
  {https://doi.org/10.1103/physrevb.93.075126} {\bibfield  {journal} {\bibinfo
  {journal} {Phys. Rev. B}\ }\textbf {\bibinfo {volume} {93}},\ \bibinfo
  {pages} {075126} (\bibinfo {year} {2016})}\BibitemShut {NoStop}%
\bibitem [{\citenamefont {Ozawa}\ \emph {et~al.}(2017)\citenamefont {Ozawa},
  \citenamefont {Taie}, \citenamefont {Ichinose},\ and\ \citenamefont
  {Takahashi}}]{ozawa17}%
  \BibitemOpen
  \bibfield  {author} {\bibinfo {author} {\bibfnamefont {H.}~\bibnamefont
  {Ozawa}}, \bibinfo {author} {\bibfnamefont {S.}~\bibnamefont {Taie}},
  \bibinfo {author} {\bibfnamefont {T.}~\bibnamefont {Ichinose}},\ and\
  \bibinfo {author} {\bibfnamefont {Y.}~\bibnamefont {Takahashi}},\ }\bibfield
  {title} {\bibinfo {title} {Interaction-driven shift and distortion of a flat
  band in an optical {Lieb} lattice},\ }\href
  {https://doi.org/10.1103/physrevlett.118.175301} {\bibfield  {journal}
  {\bibinfo  {journal} {Phys. Rev. Lett.}\ }\textbf {\bibinfo {volume} {118}},\
  \bibinfo {pages} {175301} (\bibinfo {year} {2017})}\BibitemShut {NoStop}%
\bibitem [{\citenamefont {Slot}\ \emph {et~al.}(2017)\citenamefont {Slot},
  \citenamefont {Gardenier}, \citenamefont {Jacobse}, \citenamefont {van
  Miert}, \citenamefont {Kempkes}, \citenamefont {Zevenhuizen}, \citenamefont
  {Smith}, \citenamefont {Vanmaekelbergh},\ and\ \citenamefont
  {Swart}}]{slot17}%
  \BibitemOpen
  \bibfield  {author} {\bibinfo {author} {\bibfnamefont {M.~R.}\ \bibnamefont
  {Slot}}, \bibinfo {author} {\bibfnamefont {T.~S.}\ \bibnamefont {Gardenier}},
  \bibinfo {author} {\bibfnamefont {P.~H.}\ \bibnamefont {Jacobse}}, \bibinfo
  {author} {\bibfnamefont {G.~C.}\ \bibnamefont {van Miert}}, \bibinfo {author}
  {\bibfnamefont {S.~N.}\ \bibnamefont {Kempkes}}, \bibinfo {author}
  {\bibfnamefont {S.~J.}\ \bibnamefont {Zevenhuizen}}, \bibinfo {author}
  {\bibfnamefont {C.~M.}\ \bibnamefont {Smith}}, \bibinfo {author}
  {\bibfnamefont {D.}~\bibnamefont {Vanmaekelbergh}},\ and\ \bibinfo {author}
  {\bibfnamefont {I.}~\bibnamefont {Swart}},\ }\bibfield  {title} {\bibinfo
  {title} {Experimental realization and characterization of an electronic
  {Lieb} lattice},\ }\href {https://doi.org/10.1038/nphys4105} {\bibfield
  {journal} {\bibinfo  {journal} {Nat. Phys.}\ }\textbf {\bibinfo {volume}
  {13}},\ \bibinfo {pages} {672} (\bibinfo {year} {2017})}\BibitemShut
  {NoStop}%
\bibitem [{\citenamefont {Huda}\ \emph {et~al.}(2020)\citenamefont {Huda},
  \citenamefont {Kezilebieke},\ and\ \citenamefont {Liljeroth}}]{huda20}%
  \BibitemOpen
  \bibfield  {author} {\bibinfo {author} {\bibfnamefont {M.~N.}\ \bibnamefont
  {Huda}}, \bibinfo {author} {\bibfnamefont {S.}~\bibnamefont {Kezilebieke}},\
  and\ \bibinfo {author} {\bibfnamefont {P.}~\bibnamefont {Liljeroth}},\
  }\bibfield  {title} {\bibinfo {title} {Designer flat bands in
  quasi-one-dimensional atomic lattices},\ }\href
  {https://doi.org/10.1103/PhysRevResearch.2.043426} {\bibfield  {journal}
  {\bibinfo  {journal} {Phys. Rev. Research}\ }\textbf {\bibinfo {volume}
  {2}},\ \bibinfo {pages} {043426} (\bibinfo {year} {2020})}\BibitemShut
  {NoStop}%
\bibitem [{\citenamefont {Arovas}\ \emph {et~al.}(2022)\citenamefont {Arovas},
  \citenamefont {Berg}, \citenamefont {Kivelson},\ and\ \citenamefont
  {Raghu}}]{arovas21}%
  \BibitemOpen
  \bibfield  {author} {\bibinfo {author} {\bibfnamefont {D.~P.}\ \bibnamefont
  {Arovas}}, \bibinfo {author} {\bibfnamefont {E.}~\bibnamefont {Berg}},
  \bibinfo {author} {\bibfnamefont {S.~A.}\ \bibnamefont {Kivelson}},\ and\
  \bibinfo {author} {\bibfnamefont {S.}~\bibnamefont {Raghu}},\ }\bibfield
  {title} {\bibinfo {title} {The {Hubbard} model},\ }\href
  {https://doi.org/10.1146/annurev-conmatphys-031620-102024} {\bibfield
  {journal} {\bibinfo  {journal} {Annu. Rev. Conden. Ma. P.}\ }\textbf
  {\bibinfo {volume} {13}},\ \bibinfo {pages} {239} (\bibinfo {year}
  {2022})}\BibitemShut {NoStop}%
\bibitem [{\citenamefont {Parameswaran}\ \emph {et~al.}(2013)\citenamefont
  {Parameswaran}, \citenamefont {Roy},\ and\ \citenamefont
  {Sondhi}}]{parameswaran13}%
  \BibitemOpen
  \bibfield  {author} {\bibinfo {author} {\bibfnamefont {S.~A.}\ \bibnamefont
  {Parameswaran}}, \bibinfo {author} {\bibfnamefont {R.}~\bibnamefont {Roy}},\
  and\ \bibinfo {author} {\bibfnamefont {S.~L.}\ \bibnamefont {Sondhi}},\
  }\bibfield  {title} {\bibinfo {title} {Fractional quantum hall physics in
  topological flat bands},\ }\href {https://doi.org/10.1016/j.crhy.2013.04.003}
  {\bibfield  {journal} {\bibinfo  {journal} {Cr. Phys.}\ }\textbf {\bibinfo
  {volume} {14}},\ \bibinfo {pages} {816} (\bibinfo {year} {2013})}\BibitemShut
  {NoStop}%
\bibitem [{\citenamefont {Liu}\ \emph {et~al.}(2014)\citenamefont {Liu},
  \citenamefont {Liu},\ and\ \citenamefont {Wu}}]{liu14}%
  \BibitemOpen
  \bibfield  {author} {\bibinfo {author} {\bibfnamefont {Z.}~\bibnamefont
  {Liu}}, \bibinfo {author} {\bibfnamefont {F.}~\bibnamefont {Liu}},\ and\
  \bibinfo {author} {\bibfnamefont {Y.-S.}\ \bibnamefont {Wu}},\ }\bibfield
  {title} {\bibinfo {title} {Exotic electronic states in the world of flat
  bands: {From} theory to material},\ }\href
  {https://doi.org/10.1088/1674-1056/23/7/077308} {\bibfield  {journal}
  {\bibinfo  {journal} {Chinese Phys. B}\ }\textbf {\bibinfo {volume} {23}},\
  \bibinfo {pages} {077308} (\bibinfo {year} {2014})}\BibitemShut {NoStop}%
\bibitem [{\citenamefont {Leykam}\ \emph {et~al.}(2018)\citenamefont {Leykam},
  \citenamefont {Andreanov},\ and\ \citenamefont {Flach}}]{leykam18}%
  \BibitemOpen
  \bibfield  {author} {\bibinfo {author} {\bibfnamefont {D.}~\bibnamefont
  {Leykam}}, \bibinfo {author} {\bibfnamefont {A.}~\bibnamefont {Andreanov}},\
  and\ \bibinfo {author} {\bibfnamefont {S.}~\bibnamefont {Flach}},\ }\bibfield
   {title} {\bibinfo {title} {Artificial flat band systems: {From} lattice
  models to experiments},\ }\href
  {https://doi.org/10.1080/23746149.2018.1473052} {\bibfield  {journal}
  {\bibinfo  {journal} {Adv. Phys.: X}\ }\textbf {\bibinfo {volume} {3}},\
  \bibinfo {pages} {1473052} (\bibinfo {year} {2018})}\BibitemShut {NoStop}%
\bibitem [{\citenamefont {Balents}\ \emph {et~al.}(2020)\citenamefont
  {Balents}, \citenamefont {Dean}, \citenamefont {Efetov},\ and\ \citenamefont
  {Young}}]{balents20}%
  \BibitemOpen
  \bibfield  {author} {\bibinfo {author} {\bibfnamefont {L.}~\bibnamefont
  {Balents}}, \bibinfo {author} {\bibfnamefont {C.~R.}\ \bibnamefont {Dean}},
  \bibinfo {author} {\bibfnamefont {D.~K.}\ \bibnamefont {Efetov}},\ and\
  \bibinfo {author} {\bibfnamefont {A.~F.}\ \bibnamefont {Young}},\ }\bibfield
  {title} {\bibinfo {title} {Superconductivity and strong correlations in
  {Moiré} flat bands},\ }\href {https://doi.org/10.1038/s41567-020-0906-9}
  {\bibfield  {journal} {\bibinfo  {journal} {Nat. Phys.}\ }\textbf {\bibinfo
  {volume} {16}},\ \bibinfo {pages} {725} (\bibinfo {year} {2020})}\BibitemShut
  {NoStop}%
\bibitem [{\citenamefont {Cao}\ \emph {et~al.}(2018)\citenamefont {Cao},
  \citenamefont {Fatemi}, \citenamefont {Fang}, \citenamefont {Watanabe},
  \citenamefont {Taniguchi}, \citenamefont {Kaxiras},\ and\ \citenamefont
  {Jarillo-Herrero}}]{cao18}%
  \BibitemOpen
  \bibfield  {author} {\bibinfo {author} {\bibfnamefont {Y.}~\bibnamefont
  {Cao}}, \bibinfo {author} {\bibfnamefont {V.}~\bibnamefont {Fatemi}},
  \bibinfo {author} {\bibfnamefont {S.}~\bibnamefont {Fang}}, \bibinfo {author}
  {\bibfnamefont {K.}~\bibnamefont {Watanabe}}, \bibinfo {author}
  {\bibfnamefont {T.}~\bibnamefont {Taniguchi}}, \bibinfo {author}
  {\bibfnamefont {E.}~\bibnamefont {Kaxiras}},\ and\ \bibinfo {author}
  {\bibfnamefont {P.}~\bibnamefont {Jarillo-Herrero}},\ }\bibfield  {title}
  {\bibinfo {title} {Unconventional superconductivity in magic-angle graphene
  superlattices},\ }\href@noop {} {\bibfield  {journal} {\bibinfo  {journal}
  {Nature}\ }\textbf {\bibinfo {volume} {556}},\ \bibinfo {pages} {43}
  (\bibinfo {year} {2018})}\BibitemShut {NoStop}%
\bibitem [{\citenamefont {Mizoguchi}\ and\ \citenamefont
  {Udagawa}(2019)}]{mizoguchi19}%
  \BibitemOpen
  \bibfield  {author} {\bibinfo {author} {\bibfnamefont {T.}~\bibnamefont
  {Mizoguchi}}\ and\ \bibinfo {author} {\bibfnamefont {M.}~\bibnamefont
  {Udagawa}},\ }\bibfield  {title} {\bibinfo {title} {Flat-band engineering in
  tight-binding models: {Beyond} the nearest-neighbor hopping},\ }\href
  {https://doi.org/10.1103/physrevb.99.235118} {\bibfield  {journal} {\bibinfo
  {journal} {Phys. Rev. B}\ }\textbf {\bibinfo {volume} {99}},\ \bibinfo
  {pages} {235118} (\bibinfo {year} {2019})}\BibitemShut {NoStop}%
\bibitem [{\citenamefont {Qin}\ \emph {et~al.}(2022)\citenamefont {Qin},
  \citenamefont {Schäfer}, \citenamefont {Andergassen}, \citenamefont
  {Corboz},\ and\ \citenamefont {Gull}}]{qin21}%
  \BibitemOpen
  \bibfield  {author} {\bibinfo {author} {\bibfnamefont {M.}~\bibnamefont
  {Qin}}, \bibinfo {author} {\bibfnamefont {T.}~\bibnamefont {Schäfer}},
  \bibinfo {author} {\bibfnamefont {S.}~\bibnamefont {Andergassen}}, \bibinfo
  {author} {\bibfnamefont {P.}~\bibnamefont {Corboz}},\ and\ \bibinfo {author}
  {\bibfnamefont {E.}~\bibnamefont {Gull}},\ }\bibfield  {title} {\bibinfo
  {title} {The {Hubbard} model: {A} computational perspective},\ }\href
  {https://doi.org/10.1146/annurev-conmatphys-090921-033948} {\bibfield
  {journal} {\bibinfo  {journal} {Annu. Rev. Conden. Ma. P.}\ }\textbf
  {\bibinfo {volume} {13}},\ \bibinfo {pages} {275} (\bibinfo {year}
  {2022})}\BibitemShut {NoStop}%
\bibitem [{\citenamefont {Iskin}(2021)}]{iskin21}%
  \BibitemOpen
  \bibfield  {author} {\bibinfo {author} {\bibfnamefont {M.}~\bibnamefont
  {Iskin}},\ }\bibfield  {title} {\bibinfo {title} {Two-body problem in a
  multiband lattice and the role of quantum geometry},\ }\href
  {https://doi.org/10.1103/physreva.103.053311} {\bibfield  {journal} {\bibinfo
   {journal} {Phys. Rev. A}\ }\textbf {\bibinfo {volume} {103}},\ \bibinfo
  {pages} {053311} (\bibinfo {year} {2021})}\BibitemShut {NoStop}%
\bibitem [{\citenamefont {Iskin}(2022{\natexlab{b}})}]{iskin22}%
  \BibitemOpen
  \bibfield  {author} {\bibinfo {author} {\bibfnamefont {M.}~\bibnamefont
  {Iskin}},\ }\bibfield  {title} {\bibinfo {title} {Effective-mass tensor of
  the two-body bound states and the quantum-metric tensor of the underlying
  {Bloch} states in multiband lattices},\ }\href
  {https://doi.org/10.1103/physreva.105.023312} {\bibfield  {journal} {\bibinfo
   {journal} {Phys. Rev. A}\ }\textbf {\bibinfo {volume} {105}},\ \bibinfo
  {pages} {023312} (\bibinfo {year} {2022}{\natexlab{b}})}\BibitemShut
  {NoStop}%
\bibitem [{\citenamefont {Pricoupenko}(2011)}]{pricoupenko11}%
  \BibitemOpen
  \bibfield  {author} {\bibinfo {author} {\bibfnamefont {L.}~\bibnamefont
  {Pricoupenko}},\ }\bibfield  {title} {\bibinfo {title} {Isotropic contact
  forces in arbitrary representation: Heterogeneous few-body problems and low
  dimensions},\ }\href {https://doi.org/10.1103/PhysRevA.83.062711} {\bibfield
  {journal} {\bibinfo  {journal} {Phys. Rev. A}\ }\textbf {\bibinfo {volume}
  {83}},\ \bibinfo {pages} {062711} (\bibinfo {year} {2011})}\BibitemShut
  {NoStop}%
\bibitem [{Note1()}]{Note1}%
  \BibitemOpen
  \bibinfo {note} {It is such that the binding energy $E_{2}^\protect \textrm
  {be} (\protect \mathbf {q})$ of the dimer is always defined from an unbound
  pair of a free spin-$\downarrow $ fermion plus a free spin-$\uparrow $
  fermion; the binding energy $E_{3}^\protect \textrm {be} (\protect \mathbf
  {q})$ of the trimer is defined from the dimer threshold plus a free
  spin-$\uparrow $ fermion when $E_{2}^\protect \textrm {be} (\protect \mathbf
  {q}) > 0$; the binding energy $E_{4}^\protect \textrm {be} (\protect \mathbf
  {q})$ of the tetramer is defined from the trimer threshold plus a free
  spin-$\uparrow $ fermion when $E_{3}^\protect \textrm {be} (\protect \mathbf
  {q}) > 0$, etc.}\BibitemShut {Stop}%
\bibitem [{\citenamefont {T\"orm\"a}\ \emph {et~al.}(2018)\citenamefont
  {T\"orm\"a}, \citenamefont {Liang},\ and\ \citenamefont {Peotta}}]{torma18}%
  \BibitemOpen
  \bibfield  {author} {\bibinfo {author} {\bibfnamefont {P.}~\bibnamefont
  {T\"orm\"a}}, \bibinfo {author} {\bibfnamefont {L.}~\bibnamefont {Liang}},\
  and\ \bibinfo {author} {\bibfnamefont {S.}~\bibnamefont {Peotta}},\
  }\bibfield  {title} {\bibinfo {title} {Quantum metric and effective mass of a
  two-body bound state in a flat band},\ }\href
  {https://doi.org/10.1103/PhysRevB.98.220511} {\bibfield  {journal} {\bibinfo
  {journal} {Phys. Rev. B}\ }\textbf {\bibinfo {volume} {98}},\ \bibinfo
  {pages} {220511} (\bibinfo {year} {2018})}\BibitemShut {NoStop}%
\bibitem [{\citenamefont {Huber}\ and\ \citenamefont {Altman}(2010)}]{huber10}%
  \BibitemOpen
  \bibfield  {author} {\bibinfo {author} {\bibfnamefont {S.~D.}\ \bibnamefont
  {Huber}}\ and\ \bibinfo {author} {\bibfnamefont {E.}~\bibnamefont {Altman}},\
  }\bibfield  {title} {\bibinfo {title} {Bose condensation in flat bands},\
  }\href {https://doi.org/10.1103/PhysRevB.82.184502} {\bibfield  {journal}
  {\bibinfo  {journal} {Phys. Rev. B}\ }\textbf {\bibinfo {volume} {82}},\
  \bibinfo {pages} {184502} (\bibinfo {year} {2010})}\BibitemShut {NoStop}%
\bibitem [{\citenamefont {Phillips}\ \emph {et~al.}(2015)\citenamefont
  {Phillips}, \citenamefont {De~Chiara}, \citenamefont {\"Ohberg},\ and\
  \citenamefont {Valiente}}]{phillips15}%
  \BibitemOpen
  \bibfield  {author} {\bibinfo {author} {\bibfnamefont {L.~G.}\ \bibnamefont
  {Phillips}}, \bibinfo {author} {\bibfnamefont {G.}~\bibnamefont {De~Chiara}},
  \bibinfo {author} {\bibfnamefont {P.}~\bibnamefont {\"Ohberg}},\ and\
  \bibinfo {author} {\bibfnamefont {M.}~\bibnamefont {Valiente}},\ }\bibfield
  {title} {\bibinfo {title} {Low-energy behavior of strongly interacting bosons
  on a flat-band lattice above the critical filling factor},\ }\href
  {https://doi.org/10.1103/PhysRevB.91.054103} {\bibfield  {journal} {\bibinfo
  {journal} {Phys. Rev. B}\ }\textbf {\bibinfo {volume} {91}},\ \bibinfo
  {pages} {054103} (\bibinfo {year} {2015})}\BibitemShut {NoStop}%
\bibitem [{\citenamefont {Pyykkönen}\ \emph {et~al.}(2021)\citenamefont
  {Pyykkönen}, \citenamefont {Peotta}, \citenamefont {Fabritius},
  \citenamefont {Mohan}, \citenamefont {Esslinger},\ and\ \citenamefont
  {Törmä}}]{pyykkonen21}%
  \BibitemOpen
  \bibfield  {author} {\bibinfo {author} {\bibfnamefont {V.~A.~J.}\
  \bibnamefont {Pyykkönen}}, \bibinfo {author} {\bibfnamefont
  {S.}~\bibnamefont {Peotta}}, \bibinfo {author} {\bibfnamefont
  {P.}~\bibnamefont {Fabritius}}, \bibinfo {author} {\bibfnamefont
  {J.}~\bibnamefont {Mohan}}, \bibinfo {author} {\bibfnamefont
  {T.}~\bibnamefont {Esslinger}},\ and\ \bibinfo {author} {\bibfnamefont
  {P.}~\bibnamefont {Törmä}},\ }\bibfield  {title} {\bibinfo {title}
  {Flat-band transport and {Josephson} effect through a finite-size sawtooth
  lattice},\ }\href {https://doi.org/10.1103/physrevb.103.144519} {\bibfield
  {journal} {\bibinfo  {journal} {Phys. Rev. B}\ }\textbf {\bibinfo {volume}
  {103}},\ \bibinfo {pages} {144519} (\bibinfo {year} {2021})}\BibitemShut
  {NoStop}%
\bibitem [{\citenamefont {Chan}\ \emph {et~al.}(2022)\citenamefont {Chan},
  \citenamefont {Grémaud},\ and\ \citenamefont {Batrouni}}]{chan22}%
  \BibitemOpen
  \bibfield  {author} {\bibinfo {author} {\bibfnamefont {S.~M.}\ \bibnamefont
  {Chan}}, \bibinfo {author} {\bibfnamefont {B.}~\bibnamefont {Grémaud}},\
  and\ \bibinfo {author} {\bibfnamefont {G.~G.}\ \bibnamefont {Batrouni}},\
  }\bibfield  {title} {\bibinfo {title} {Pairing and superconductivity in
  quasi-one-dimensional flat-band systems: {Creutz} and sawtooth lattices},\
  }\href {https://doi.org/10.1103/physrevb.105.024502} {\bibfield  {journal}
  {\bibinfo  {journal} {Phys. Rev. B}\ }\textbf {\bibinfo {volume} {105}},\
  \bibinfo {pages} {024502} (\bibinfo {year} {2022})}\BibitemShut {NoStop}%
\bibitem [{\citenamefont {White}(1992)}]{srwhite}%
  \BibitemOpen
  \bibfield  {author} {\bibinfo {author} {\bibfnamefont {S.~R.}\ \bibnamefont
  {White}},\ }\bibfield  {title} {\bibinfo {title} {Density matrix formulation
  for quantum renormalization groups},\ }\href
  {https://doi.org/10.1103/PhysRevLett.69.2863} {\bibfield  {journal} {\bibinfo
   {journal} {Phys. Rev. Lett.}\ }\textbf {\bibinfo {volume} {69}},\ \bibinfo
  {pages} {2863} (\bibinfo {year} {1992})}\BibitemShut {NoStop}%
\bibitem [{\citenamefont {Schollw\"{o}ck}(2011)}]{schollwock11}%
  \BibitemOpen
  \bibfield  {author} {\bibinfo {author} {\bibfnamefont {U.}~\bibnamefont
  {Schollw\"{o}ck}},\ }\bibfield  {title} {\bibinfo {title} {The density-matrix
  renormalization group in the age of matrix product states},\ }\href
  {https://doi.org/10.1016/j.aop.2010.09.012} {\bibfield  {journal} {\bibinfo
  {journal} {Ann. Phys.}\ }\textbf {\bibinfo {volume} {326}},\ \bibinfo {pages}
  {96} (\bibinfo {year} {2011})}\BibitemShut {NoStop}%
\bibitem [{\citenamefont {Fishman}\ \emph {et~al.}(2020)\citenamefont
  {Fishman}, \citenamefont {White},\ and\ \citenamefont
  {Stoudenmire}}]{itensor}%
  \BibitemOpen
  \bibfield  {author} {\bibinfo {author} {\bibfnamefont {M.}~\bibnamefont
  {Fishman}}, \bibinfo {author} {\bibfnamefont {S.~R.}\ \bibnamefont {White}},\
  and\ \bibinfo {author} {\bibfnamefont {E.~M.}\ \bibnamefont {Stoudenmire}},\
  }\href@noop {} {\bibinfo {title} {The \mbox{ITensor} software library for
  tensor network calculations}} (\bibinfo {year} {2020}),\ \Eprint
  {https://arxiv.org/abs/2007.14822} {arXiv:2007.14822} \BibitemShut {NoStop}%
\bibitem [{\citenamefont {Weinberg}\ and\ \citenamefont {Bukov}(2019)}]{ed}%
  \BibitemOpen
  \bibfield  {author} {\bibinfo {author} {\bibfnamefont {P.}~\bibnamefont
  {Weinberg}}\ and\ \bibinfo {author} {\bibfnamefont {M.}~\bibnamefont
  {Bukov}},\ }\bibfield  {title} {\bibinfo {title} {{QuSpin: a Python Package
  for Dynamics and Exact Diagonalisation of Quantum Many Body Systems. Part II:
  bosons, fermions and higher spins}},\ }\href
  {https://doi.org/10.21468/SciPostPhys.7.2.020} {\bibfield  {journal}
  {\bibinfo  {journal} {SciPost Phys.}\ }\textbf {\bibinfo {volume} {7}},\
  \bibinfo {pages} {20} (\bibinfo {year} {2019})}\BibitemShut {NoStop}%
\bibitem [{\citenamefont {Valiente}\ \emph {et~al.}(2010)\citenamefont
  {Valiente}, \citenamefont {Petrosyan},\ and\ \citenamefont
  {Saenz}}]{valiente10}%
  \BibitemOpen
  \bibfield  {author} {\bibinfo {author} {\bibfnamefont {M.}~\bibnamefont
  {Valiente}}, \bibinfo {author} {\bibfnamefont {D.}~\bibnamefont
  {Petrosyan}},\ and\ \bibinfo {author} {\bibfnamefont {A.}~\bibnamefont
  {Saenz}},\ }\bibfield  {title} {\bibinfo {title} {Three-body bound states in
  a lattice},\ }\href {https://doi.org/10.1103/PhysRevA.81.011601} {\bibfield
  {journal} {\bibinfo  {journal} {Phys. Rev. A}\ }\textbf {\bibinfo {volume}
  {81}},\ \bibinfo {pages} {011601} (\bibinfo {year} {2010})}\BibitemShut
  {NoStop}%
\bibitem [{\citenamefont {Iskin}\ and\ \citenamefont
  {Keleş}(2022)}]{iskin22d}%
  \BibitemOpen
  \bibfield  {author} {\bibinfo {author} {\bibfnamefont {M.}~\bibnamefont
  {Iskin}}\ and\ \bibinfo {author} {\bibfnamefont {A.}~\bibnamefont {Keleş}},\
  }\href@noop {} {\bibinfo {title} {Dimers, trimers, tetramers, and other
  multimers in a multiband bose-hubbard model}} (\bibinfo {year} {2022}),\
  \Eprint {https://arxiv.org/abs/2208.01429} {arXiv:2208.01429} \BibitemShut
  {NoStop}%
\end{thebibliography}%
\end{document}